\documentclass[useAMS,usenatbib]{mn2e}

\newcommand{\ii}{\'{\i}}

\newcommand{\uu}{\'u}
\newcommand{\ee}{\'e}
\def\lsim{\ ^{<}\!\!\!\!_{\sim}\>}
\def\gsim{\ ^{>}\!\!\!\!_{\sim}\>}

\usepackage[dvips]{graphicx}

\title[Cometary masses derived from non-gravitational forces]{Cometary masses derived from non-gravitational forces}

\author[Andrea Sosa and Julio A. Fern\'andez]{Andrea Sosa$^{1}$\thanks{E-mail:
asosa@fisica.edu.uy (AS); julio@fisica.edu.uy (JAF)} and Julio A. Fern\'andez$^{1}$\footnotemark[1]\\
$^{1}$Departamento de Astronom\ii a, Facultad de Ciencias, Igu\'a 4225, Montevideo, 11400, Uruguay}

\begin{document}

%\date{Accepted 1988 December 15. Received 1988 December 14; in original form 1988 October 11}

\pagerange{\pageref{firstpage}--\pageref{lastpage}} \pubyear{2008}

\maketitle

\label{firstpage}

\begin{abstract}
We compute masses and densities for ten periodic comets with known sizes: 1P/Halley, 2P/Encke, 6P/d'Arrest, 9P/Tempel 1, 10P/Tempel 2, 19P/Borrelly, 22P/Kopff, 46P/Wirtanen, 67P/Churyumov-Gerasimenko and 81P/Wild 2.
The method follows the one developed by Rickman and colleagues (Rickman 1986, 1989; Rickman {\em et al.} 1987), which is  based on the gas production curve and on the change in the orbital period due to the non-gravitational force. The gas production curve is inferred from the visual lightcurve. We found that the computed masses cover more than three orders of magnitude: $\simeq$ (0.3 - 400) $\times$ 10$^{12}$ kg. The computed densities are in all cases very low ($\lsim$ 0.8 g cm$^{-3}$), with an average value of 0.4 g cm$^{-3}$, in agreement with previous results and models of the cometary nucleus depicting it as a very porous object. The computed comet densities turn out to be the lowest among the different populations of solar system minor bodies, in particular as compared to those of near-Earth asteroids (NEAs). We conclude that the model applied in this work, in spite of its simplicity (as compared to more sophisticated thermophysical models applied to very few comets), is useful for a statistical approach to the mean density of the cometary nuclei. However, we cannot assess from this simple model if there is a real dispersion among the bulk densities of comets that could tell us about differences in physical structure (porosity) and/or chemical composition.
\end{abstract}


\section{Introduction}

Comets have a great cosmogonical interest since they are thought to be the relics
of planet formation. They are ice-rich bodies formed beyond
the {\it snowline}, i.e. the distance from the Sun at which water ice could
condense. Comet material is presumably the most pristine so far
known in the solar system that may keep the record of the grain accretion
process in the protoplanetary disk. Micrometer-sized grains will tend to collect after gentle collisions
into loose aggregates. Since comets are very small bodies, their
self-gravity was never large enough to compact the material, so they are
expected to conserve their primordial porous structure. Therefore, the comet
nucleus is regarded as a very fragile and low-density structure. Donn (1963)
estimated that the mixture of ices and meteoric matter would form an aggregate
with a density of a few tenths g cm$^{-3}$. Later Weissman (1986) proposed
a nuclear structure consisting of aggregates of various sizes that he called
the 'primordial rubble-pile' model.

The comet bulk density is a key parameter to learn about the geochemistry and the degree of
compactness of the
comet material. Masses and sizes of comets are necessary to derive their bulk
densities. The determination of these two physical parameters is not trivial.
The main difficulty to determine the comet size is that comets are usually
shrouded by a coma of dust and gas. Nevertheless, great efforts have been made
to observe Jupiter family comets near aphelion where they are inactive or have
only residual activity, so estimates of comet sizes have been possible for a
large sample of Jupiter family comets (e.g. Fern\'andez {\em et al.} 1999, Tancredi
{\em et al.} 2000, 2006). It has also been possible to obtain very accurate sizes of a few comets from close up images taken from spacecraft during flyby encounters. These are the cases of 1P/Halley (e.g. Keller {\em et al.} 1987), 19P/Borrelly (e.g. Soderblom {\em et al.} 2002), 81P/Wild 2 (e.g. Brownlee {\em et al.} 2004, Duxbury {\em et al.} 2004, Howington-Kraus {\em et al.} 2005), and 9P/Tempel 1 (e.g. A'Hearn {\em et al.} 2005).

The determination of comet masses has been an even more difficult task.
In 1805 Laplace could set an upper limit of 1/5000 of the Earth mass for the mass of comet Lexell by noting that the Earth's motion did not suffer any perceptible perturbation after a close encounter with the comet in 1770. From then new attempts to derive comet masses led to even smaller masses showing that comets were indeed very small bodies.   Vorontsov-Velyaminov (1946) derived minimum initial masses of $10^{18}$ g for comet 109P/Swift-Tuttle and of $10^{16}$ g for 55P/Tempel-Tuttle, based on the mass estimates for the associated Perseids and Leonids meteor streams respectively, and assuming that the masses of the meteor streams were only a fraction of the masses still remaining in the
respective nuclei. Asphaug and Benz (1996) modeled the tidal disruption of comet Shoemaker-Levy 9, and the formation of a long chain of clumps of fragments, during its near-miss with Jupiter in 1992. The authors assumed a stengthless body. By comparing the model results with observations, they obtained as a best-fit solution a diameter $\sim$ 1.5 km and a density of $\sim$ 0.6 g cm$^{-3}$ for SW9.

All the results mentioned before can only be applied to particular cases and are very uncertain. Since we do not know about binary comets or comet satellites, and no spacecraft has measured the gravitational field of a comet (except for the {\em Deep Impact} experiment with comet 9P/Tempel 1, as commented in section 6), the only general effect that can give us a direct estimate of the mass of the comet nucleus is the non-gravitational acceleration as shall be analyzed in the next section. A method for computing cometary masses based on the non-gravitational effect was introduced by Rickman (1986, 1989), Rickman {\em et al.} (1987), and a somewhat similar approach was presented by Sagdeev {\em et al.} (1987).

In this paper we plan to reevaluate the suitability of the non-gravitational force as a tool to estimate cometary masses following a procedure somewhat similar to that developed by Rickman and colleagues. We will also analyze how the uncertainties involved in the different physical parameters related to the problem affect the final results. Finally, we will take advantage of the progress made in the determination of cometary sizes during the last decade to compute bulk densities from the derived non-gravitational masses.

Section 2 presents a brief description of the non-gravitational forces. The method used, the model parameters, and the data of the selected comets for our study are summarized in Section 3. In Section 4 we discuss the problems in the cometary photomety and the empirical correlation between visual magnitudes and gas production rates. In Section 5 we present and discuss the visual lightcurves obtained for our sample of comets. The results are presented and discussed in Section 6, with a further analysis in Section 7. Finally, a summary with our main conclusions is made in Section 8.

\section{Non-gravitational forces}

The main non-gravitational effect that can be detected in a periodic comet is a change in its orbital period, with respect to that derived from purely gravitational theory. Whipple (1950) proposed an icy conglomerate model for the cometary nuclei, and showed that the momentum transferred to the nucleus by the outgassing could cause the observed non-gravitational effect, an idea that was already advanced more than a century before by Bessel (1836). Because of the thermal inertia and the nucleus's rotation, Whipple showed that the direction of the net outgassing will be deviated a certain angle (known as the {\em lag angle}) with respect to the subsolar point. Today we know that some geometrical factors like the shape, the topography, the spin orientation or the location of the active areas on the nucleus surface, are also crucial in the determination of the lag angle, as suggested by thermophysical models (Davidsson and Guti\ee rrez 2004, 2005, 2006, and Davidsson, Guti\ee rrez and Rickman 2007). We emphasized that in this work the term 'non-gravitational forces'  means the momentum transferred to the nucleus by the outgassing, since other non-gravitational forces acting on minor bodies of the Solar System are negligible in the case of comets.

The change $\Delta P$ in the orbital period $P$ due to the non-gravitational force can be expressed in terms of the radial $J_r$ and transverse $J_t$ components of the non-gravitational acceleration $J$ by means of the Lagrangian planetary equations under the Gaussian form, which leads to

\begin{center}
\begin{equation}
\Delta P \  =  \ \frac{6 \pi \sqrt{1-e^2}}{n^2} \bigg[ \frac{e}{a(1-e^2)}  \int_{0}^{P} J_r \sin(f) dt + \int_{0}^{P}\frac{J_t}{r} dt \bigg],
\label{deltap}
\end{equation}
\end{center}

\noindent
where $n$ is the mean motion, $a$ the semimajor axis, $e$ the eccentricity, $f$ the true anomaly, and $r$ the heliocentric distance. The direction of the non-gravitational acceleration (which is in the opposite direction to that of the net outgassing) can be described by the angle $\eta$  with respect to the antisolar direction (i.e. the lag angle), and an azimuthal angle $\phi$ in the plane perpendicular to the antisolar direction. The components of the non-gravitational acceleration can be expressed in terms of the angles $\eta$ and $\phi$ as follows
%\begin{center}
\begin{eqnarray}
J_r  &=&  J \cos(\eta),  {} \nonumber\\
J_t  &=& J \sin(\eta) \cos(\phi), {} \nonumber\\
J_n  &=&  J \sin(\eta) \sin(\phi)
\label{componentes}
\end{eqnarray}
%\end{center}

Equation (\ref{deltap}) shows that $\Delta P$ depends on the radial and transverse components $J_r$ and $J_t$, but not on the normal component $J_n$ of the non-gravitational acceleration (i.e. the component perpendicular to the orbital plane).

\indent
We find some confusion in the literature as regards to what we understand by $\Delta P$. Some authors refer to a delay or advance in the time of the perihelion passage, instead of the change in the orbital period. These quantities are not equivalent since the change in the time of the perihelion passage is determined not only by the change in the period $\Delta P$, but also by the change in the angular orbital parameters.

Finally, according to Whipple (1950), we can relate the non-gravitational acceleration $J$ and the comet's mass $M$ by means of the conservation of momentum i.e.

\begin{center}
\begin{equation}
M\vec{J} \ = \ -Q m \vec{u},
\label{momento}
\end{equation}
\end{center}

\noindent
where $Q$ is the gas production rate, $\vec{u}$ is the effective outflow velocity, and $m$ is the average molecular mass.
%The effective outflow velocity can be expressed as $\vec{u}$ = $\zeta \vec{v}_{th}$, where $\zeta$ is the momentum transfer efficiency (by unit mass) and $\vec{v}_{th}$ is the mean thermal velocity of the gas molecules.

\section{The method}

In this section we introduce the assumptions we made in order to simplify the problem of the mass determination.

By substituting $J_r$ and $J_t$ of eq. (\ref{deltap}) by the expressions given by eqs. (\ref{componentes}) and (\ref{momento}), and then solving for the mass, we obtain the following expression

\begin{center}
\begin{eqnarray}
M \ &=& \ \frac{6 \pi \sqrt{1-e^2}}{n^2} \bigg[ \frac{e}{a(1-e^2)} \int_{0}^{P} \frac{Q m u}{\Delta P} \cos(\eta) \sin(f) dt + {} \nonumber\\
& & \int_{0}^{P}\frac{Q m u}{\Delta P}\frac{1}{r} \sin(\eta) \cos(\phi) dt \bigg].
\label{masa-exacta}
\end{eqnarray}
\end{center}

In order to solve the integrals in eq. (\ref{masa-exacta}) we should know the outflow velocity and the angles as functions of time or the heliocentric distance. Our first approach will be to consider an average value $<u>$ for all escaping molecules during an orbital revolution of period $P$. We also consider average values for the angles  $<\cos(\eta)>$ and $<\sin(\eta) \cos(\phi)>$. We assume these values as constants for all the comets of our sample. We also assume that water, as the principal volatile component in the cometary composition, dominates the gas production in the inner solar system (say $r$ $\lsim$ 3 AU), so we take $m$ as the water molecular mass. Therefore, we can express eq. (\ref{masa-exacta}) as follows

\begin{center}
\begin{equation}
M \ \cong \ \frac{6 \pi \sqrt{1-e^2}}{n^2 \Delta P} m <u> (I_r \ + \ I_t),
\label{masa-aprox}
\end{equation}
\end{center}

\noindent where $I_r$ and $I_t$ are given by

\begin{center}
\begin{displaymath}
I_r  \ = \  \frac{e}{a(1-e^2)}<\cos(\eta)>\int_{0}^{P} Q \sin(f) dt,
\end{displaymath}
\begin{displaymath}
I_t  \ = \ <\sin(\eta) \cos(\phi)>\int_{0}^{P}\frac{Q}{r} dt
\end{displaymath}
\end{center}

The computation of the effective outflow velocity $<u>$ relies on sophisticated thermophysical modeling of the exchange of linear momentum between the sublimating gas and the nucleus surface. The value of $<u>$ depends not only on nucleus temperature, but also on shape and location of active areas, therefore is highly uncertain. As an educated guess, we consider $<u>$ = $\zeta v_{th}$, where $v_{th}$ is the mean thermal velocity (in an equilibrium gas), and $\zeta$ is a dimensionless quantity that contains any difference between the physically meaningful parameter $<u>$ and the arbitrary chosen normalization quantity $v_{th}$. The mean thermal velocity is given by $v_{th}$ = $(8kT/\pi m)^{1/2}$, where $k$ is Boltzmann's constant, $T$ is the temperature of the sublimating gases (which is about 200 K for the range of heliocentric distances of interest $\sim$ 0.5 - 3 AU), and $m$, as before, the water molecular mass.  For $\zeta$ $\simeq$ 0.5 - 0.6 we obtain the value $< u >$ $\simeq$ 0.25 km s$^{-1}$ (Wallis and Macpherson 1981, Rickman 1989).  Some other works point to a somewhat higher value of $<u>$ (around 0.3 km/s) (Peale 1989, Davidsson and Guti\ee rrez 2004). Taking into account the range of published values, we finally adopted the mean value of (0.27 $\pm$ 0.1) km s$^{-1}$ for the effective outflow velocity.

As we said before, in general $\eta$ and $\phi$ will vary with the orbital position. Unfortunately, we cannot know the values of these angles, unless we can study the comet nucleus {\em in situ}. Therefore, we depend on educated guesses and averages to sort this problem out. For most reasonable combinations of tentative values for the rotational period, the thermal inertia of the outermost layers of the nucleus and the heliocentric distance, Rickman (1986) and Rickman {\em et al.} (1987) obtained, by means of thermal models, $\eta$ $\leq$ 30$^{\circ}$. In more recent works with thermophysical models Davidsson and Guti\ee rrez (2004, 2005, 2006) also found small values for $\eta$. These results confirm our guess that the net non-gravitational force may be deviated from the radial direction just by a small angle.

So we have only positive values for the parameter $<\cos(\eta)>$, while the parameter $<\sin(\eta)\cos(\phi)>$ has not such a constraint. This leads to an indetermination in the sign of the transverse non-gravitational contribution, as defined in eq. (\ref{masa-aprox}). To sort this problem out, we will use the following approach

\begin{center}
\begin{displaymath}
<\sin(\eta)\cos(\phi)> \cong sig(\Delta P) <|\sin(\eta)\cos(\phi)|>.
\end{displaymath}
\end{center}

\noindent
To explain this assumption, we refer to eq. (\ref{masa-aprox}). There we can see that, for a comet with a symmetric curve $Q$(t) with respect to perihelion (this implies a symmetric lightcurve, as assumed in this work) the first integral will vanish,  since a symmetric Q will also imply a symmetric $J_r$, so the change $\Delta P$ will depend only on the transverse component $J_t$. In this case it is easy to see that $\Delta  P$ and $J_t$ must have the same sign, and also $\Delta P$ and $<\sin(\eta) \cos(\phi)>$.

Yet, as pointed out by Rickman (1986), most comet lightcurves are moderately or highly asymmetric (see for instance the lightcurves in Section 5), so the integral of the term containing $J_r$ will not vanish, and in some cases may become dominant. When one of the terms defined in eq. (\ref{masa-aprox}) clearly dominates the other, there is no ambiguity in the result: (i) when $I_r$ $>>$ $I_t$, the uncertainty in the sign of the transverse term contributes to the overall uncertainty in the mass as any other source of error, and (ii) when $I_r$ $<<$ $I_t$, there is no ambiguity as explained before. The problem of the indetermination in the sign of the transverse term arises when both terms have similar absolute values, in which case we cannot easily discern if $I_t$ adds or subtracts to $I_r$. Another problem related to the case where $I_r$ $\approx$ $I_t$ arises because the errors may make the difference $I_r$ - $I_t$ either positive or negative, leading for one of the signs to an unrealistic value $M$ $<$ 0.

If we further assume a uniform random distribution in the range [0, 2$\pi$] for the angle $\phi$, we finally get the nominal values for our model parameters

 \begin{center}
 \begin{displaymath}
 <\cos(\eta)> \approx 1, \ \ <|\sin(\eta)\cos(\phi)|> \approx 0.1
 \end{displaymath}
 \end{center}

Now we turn our attention to the remaining parameters or quantities involved in the computation of the nucleus mass: the gas production curve, the relevant orbital parameters (semimajor axis and eccentricity), and the non-gravitational effect $\Delta P$. Unlike the parameters discussed before, these parameters are specific for each comet.

As we can see from eq. (\ref{masa-aprox}), the computation of the cometary mass requires to know from observations the shape of the curve $Q(t)$. There are only a few cases for which we have good measurements of gas production rates at the different orbital positions, so in general we have to rely on lightcurves, and hence, assume some formula to approximately convert total visual heliocentric magnitudes $m_h$ (i.e. the apparent magnitudes $m$ corrected for the geocentric distance $\Delta$: $m_h \ = \ m - 5 \log\Delta$)
 to water production rates. In this regard we use the following empirical law, introduced by Festou (1986)

\begin{equation}
\log Q \ = a_1 \times m_h + a_2
\label{calibracion}
\end{equation}

\noindent
We adopt for the coefficients $a_1$ and $a_2$ the values found by Jorda {\em et al.} (2008)  (see Section 4.2 for a discussion of the empirical correlation between total visual magnitudes and gas production rates). In Table 1 we summarize the values adopted for the model parameters $m$, $<u>$, $<|\cos(\eta)|>$, $<|\sin(\eta)\cos(\phi)|>$, $a_1$ and $a_2$.

The orbital elements of the selected comets are shown in Table 2. For the orbital parameters $a$ and $e$ we take the values corresponding to the epoch closer to the perihelion passage. The symbols represent: $\tau$ the time of the perihelion passage, $q$ the perihelion distance (AU), $e$ the eccentricity, $P$ the orbital period (years), $\omega$ the perihelion argument ($^\circ$), $\Omega$ the longitude of the ascending node ($^\circ$), and $i$ the inclination ($^\circ$). The orbital elements and the epoch are referred to the 2000.0 equinox. The elements were obtained from the catalogue of cometary orbits of Marsden and Williams (2008). For those comets for which the lightcurves were based on photometric data corresponding to more than one apparition, an average of the values for each epoch was made. The mean motion $n$ and the orbital period $P$ were computed from the value used for the parameter $a$.

The non-gravitational effect $\Delta P$ was computed from the non-gravitational orbital solution for the comet. In this case the equation of motion has three non-gravitational parameters $A_1$, $A_2$, $A_3$ for the radial, transverse and normal components of the non-gravitational acceleration. The change $\Delta P$ will be independent of $A_3$ and, for the symmetric case, also of $A_1$, so it will only depend on $A_2$ through the equation (Marsden {\em et al.} 1973, Festou {\em et al.} 1990):

 \begin{center}
\begin{equation}
\Delta P \ = \ \frac{6 \pi \sqrt{1-e^2}}{n^2} A_2 \int_{0}^{P} \frac{g(r)}{r} dt
\label{marsden}
\end{equation}
\end{center}

  The computed values of $A_2$ are shown in Table 3. Although the model used by Marsden {\em et al.} (1973) (known as the {\em standard symmetric model}) was questioned in a physical sense to be not too realistic (since it assumes a symmetric outgassing curve with respect to perihelion), it has achieved in practice good fits to the astrometric observations of a large number of comets.

\noindent
  The parameter $A_2$ represents the transverse component of the non-gravitational acceleration at 1 AU from the Sun, and $g(r)$ is an empirical function which describes the variation of the water snow sublimation rate with respect to the heliocentric distance

\begin{center}
\begin{equation}
g(r) \ = \ \alpha \bigg(\frac{r}{r_0}\bigg)^{-m} \bigg[1 + \bigg(\frac{r}{r_0}\bigg)^{n} \bigg]^{-k},
\label{gr}
\end{equation}
\end{center}

\noindent
  where $\alpha$ = 0.1113, $m$ = 2.15, $n$ = 5.093, $k$ = 4.6142, and $r_0$ = 2.808 AU, (Marsden {\em et al.} 1973).

It may be argued that there may be a contradiction between the derivation of $\Delta P$ from the symmetric model and the use of a more realistic asymmetric model. The explanation of this seemingly contradictory procedure is that $A_1$, $A_2$ arise from the best-fit orbital solution from Marsden and Williams (2008) which is based on the symmetric model developed by Marsden {\em et al.} (1973). Therefore, for consistency $\Delta P$ has to be obtained from the $A_2$ value for the symmetric model. We expect that $\Delta P$ values derived from asymmetric models should lead to a reasonable agreement with the values derived from eq. (\ref{marsden}), as far as we get good-quality orbit solutions for the comet's astrometric positions based either on a symmetric or an asymmetric model. In order to check this assumption we also derived $\Delta P$ from Yeomans and Chodas's (1989) asymmetric model. In this regard the JPL Small-Body Database Browser (http://ssd.jpl.nasa.gov/sbdb.cgi) provides sets ($A_1$, $A_1$, $A_1$, $DT$) for a few Jupiter family comets derived from the latter model. $DT$ represents the the time offset of the non-gravitational force maximum from the time of the perihelion passage. From the JPL web site we chose comets 6P, 10P, 46P, 67P and 81P. The results (shown in Table 4) indicate that the values derived from both models are consistent, as expected.

\begin{table*}
 \centering
 \begin{minipage}{140mm}
\begin{footnotesize}
\caption{Summary of model parameters.}
\begin{tabular}{l l l l} \hline
Description & Symbol & Nominal value & Unit\\ \hline
Water molecular mass & $m$ & 2.98897 $\times$ 10$^{-26}$& kg\\
Effective outflow velocity (averaged value)& $< u >$ & 0.27 & km s$^{-1}$ \\
Angular parameter & $<\cos(\eta)>$ & 1.0 & \\
Angular parameter & $<|\sin(\eta)\cos(\phi)|>$ & 0.1 & \\
Calibration coefficient  & $a_1$ & -0.2453 &\\
Calibration coefficient  & $a_2$ &30.675 &\\
\hline
\end{tabular}
\end{footnotesize}
\end{minipage}
\end{table*}

\begin{table*}
 \centering
 \begin{minipage}{140mm}
\begin{footnotesize}
\caption{Orbital parameters of the studied comets.}
\begin{tabular}{l l l l l l l l l} \hline
Comet& $\tau$ &  $q$ & $e$ & $P$ & $\omega$ & $\Omega$ & $i$ & Epoch \\ \hline
 1P& 1986 Feb. 9.4589&  0.587104& 0.967277& 76.0& 111.8657& 58.8601& 162.2422& 86 Feb. 19\\ \hline
 2P& 1990 Oct. 28.5678& 0.330890& 0.850220& 3.28& 186.2348& 334.7492& 11.9450& 90 Nov. 5\\
   & 1994 Feb.  9.4785& 0.330918& 0.850211& 3.28& 186.2723& 334.7273& 11.9401& 94 Feb. 17 \\
   & 1997 May  23.5979& 0.331397& 0.850013& 3.28& 186.2733& 334.7202& 11.9293& 97 June 1 \\
   & 2000 Sept. 9.6669& 0.339539& 0.846898& 3.30& 186.4842& 334.5993& 11.7554& 00 Sept.13\\ \hline
 6P& 1995 July 27.3234& 1.345813& 0.614042& 6.51& 178.0502& 138.9895& 19.5239& 95 July 22\\ \hline
 9P& 1994 July  3.3141& 1.494151& 0.520255& 5.50& 178.9009&  68.9864& 10.5519& 94 June 17\\
   & 2005 July  5.3145& 1.506166& 0.517491& 5.52& 178.8382&  68.9380& 10.5301& 05 July 9\\ \hline
10P& 1983 June  1.5367& 1.381401& 0.544895& 5.29& 190.9473& 119.8323& 12.4338& 83 May  26\\
   & 1988 Sept.16.7345& 1.383426& 0.544429& 5.29& 191.0643& 119.7925& 12.4282& 88 Oct.  6\\
   & 1999 Sept. 8.4214& 1.481678& 0.522818& 5.47& 195.0232& 118.2114& 11.9767& 99 Sept.19\\ \hline
19P& 1987 Dec. 18.3239& 1.356795& 0.624213& 6.86& 353.3370&  75.4332& 30.3254& 87 Dec. 31\\
   & 1994 Nov.  1.4946& 1.365126& 0.622800& 6.88& 353.3587&  75.4238& 30.2708& 94 Oct. 15\\
   & 2001 Sept.14.7306& 1.358204& 0.623891& 6.86& 353.3750&  75.4249& 30.3248& 01 Sept. 8\\ \hline
22P& 1983 Aug. 10.3011& 1.576327& 0.544526& 6.44& 162.8877& 120.9308&  4.7210& 83 Aug. 14\\
   & 1996 July  2.1914& 1.579573& 0.544069& 6.45& 162.8377& 120.9065&  4.7211& 96 July 16\\ \hline
45P& 1990 Sept.12.6843& 0.541252& 0.821869& 5.30& 325.7863&  89.3080&  4.2198& 90 Sept.26\\
   & 1995 Dec. 25.9844& 0.532050& 0.824231& 5.27& 326.0532&  89.1561&  4.2484& 95 Dec. 29\\
   & 2001 Mar. 29.9269& 0.528413& 0.825080& 5.25& 326.1311&  89.0828&  4.2556& 01 Apr.  1\\ \hline
46P& 1991 Sept.20.6255& 1.083306& 0.652242& 5.50& 356.1687&  82.2930& 11.6820& 91 Sept.21\\
   & 1997 Mar. 14.1499& 1.063769& 0.656748& 5.46& 356.3418&  82.2051& 11.7226& 97 Mar. 13\\ \hline
67P& 1982 Nov. 12.0993& 1.306144& 0.629152& 6.61&  11.3667&  51.0145&  7.1170& 82 Nov.  7\\
   & 1996 Jan. 17.6560& 1.300032& 0.630193& 6.59&  11.3861&  51.0070&  7.1135& 95 Dec. 29\\ \hline
81P& 1990 Dec. 16.9164& 1.578061& 0.540980& 6.37&  41.6421&  136.2048& 3.2437& 90 Dec. 15\\
   & 1997 May   6.6277& 1.582623& 0.540220& 6.39&  41.7681&  136.1564& 3.2424& 97 Apr. 22\\
   & 2003 Sept.25.9302& 1.590367& 0.538787& 6.40&  41.7516&  136.1413& 3.2400& 03 Oct.  8\\ \hline

\end{tabular}
\end{footnotesize}
\end{minipage}
\end{table*}

\begin{table*}
 \centering
 \begin{minipage}{84mm}
\begin{footnotesize}
\caption{Non-gravitational effect in the studied comets.}
\begin{tabular}{l l l l} \hline
Comet & Apparition & $A_2$ $\times$10$^{8}$ \footnote{Source: Marsden and Williams (2008).} & $\Delta P$\\
             &                    & (AU/days$^2$) & (days)\\ \hline
1P/Halley & 1986 & +0.0155 & +4.1063 \\ \hline
2P/Encke & 1990 & -0.0016 & -0.0048 \\
         & 1994 & -0.0007 & -0.0021 \\
         & 1997 & -0.0007 & -0.0021\\
         & 2000 & -0.0007 & -0.0021 \\ \hline
6P/d'Arrest & 1995 & +0.0993  & +0.1297  \\ \hline
9P/Tempel 1  & 1994 & +0.0017 & +0.0013  \\
            & 2005 & +0.0017 & +0.0013  \\ \hline
10P/Tempel 2 & 1983 & +0.0014  & +0.0013  \\
             & 1988 & +0.0014  & +0.0013  \\
	     & 1999 & +0.0014 & +0.0011 \\ \hline
19P/Borrelly & 1987 & -0.0376 & -0.0522 \\
             & 1994 & -0.0473 & -0.0651 \\
             & 2001 & -0.0473 & -0.0655 \\ \hline
22P/Kopff & 1983 & -0.1127 & -0.0962 \\
          & 1996 & -0.1073 & -0.0912 \\ \hline
45P/H-M-P & 1990 & -0.0553 & -0.2042  \\
          & 1995 & -0.0505 & -0.1887  \\
          & 2001 & -0.0505 & -0.1893  \\ \hline
46P/Wirtanen & 1991 & -0.1673 & -0.2557  \\
             & 1997 & -0.1373 & -0.2137  \\ \hline
67P/C-G & 1982 & +0.0096 & +0.0137 \\
        & 1996 & +0.0096 & +0.0138 \\ \hline
81P/Wild 2 & 1990 & +0.0092 & +0.0077 \\
           & 1997 & +0.0092 & +0.0077 \\
           & 2003 & +0.0092 & +0.0076 \\ \hline
\end{tabular}
\end{footnotesize}
\end{minipage}
\end{table*}

\begin{table*}
 \centering
 \begin{minipage}{140mm}
\begin{footnotesize}
\caption{Non-gravitational effect according to two different models.}
\begin{tabular}{l c c c c c c c c c} \hline

Comet & \multicolumn{3}{|c}{Symmetrical model} & & \multicolumn{5}{|c}{Asymmetrical model} \\
 & \multicolumn{3}{|c}{(Marsden {\em{et al.}} 1973)}& & \multicolumn{5}{|c}{(Yeomans \& Chodas 1989)} \\ \cline{2-4} \cline{6-10}
 & Apparition & Parameters\footnote{Source: Marsden and Williams (2008).} & $\Delta P$ & & Apparition & \multicolumn{3}{c}{Parameters\footnote{Source: JPL Small-Body Database Browser.}} & $\Delta P$ \\ \cline{3-3} \cline{7-9}
 & & & & & & & & & \\
 & & $A_2$ $\times$10$^{8}$   & & & & $A_1\times10^{8}$ & $A_2\times10^{8}$ & $DT$ &   \\
 & & (AU/days$^2$) &(days) & & & \multicolumn{2}{c}{(AU/days$^2$)}& (days) & (days) \\ \hline
6P/d'Arrest & 2008 & +0.0993  & +0.1288 & &2008 &+0.3804 &-0.0117 &+97.3 & +0.1302\\
10P/Tempel 2 & 2005 & +0.0014  & +0.0012 & &2006 & +0.0325& -0.0010&+24.3 &+0.0011 \\
46P/Wirtanen & 2008 & -0.1373 & -0.2148 & & 2007& +0.3735 &-0.1302 & -4.4&-0.2159 \\
67P/C-G & 2009 & +0.0096 & +0.0145 & & 2008& +0.1160& -0.0062&+38.8 & +0.0171\\
81P/Wild 2 & 2003 & +0.0092 & +0.0076 & & 2006& +0.1489& +0.0333& -60.6&+0.0078\\ \hline
\end{tabular}
\end{footnotesize}
\end{minipage}
\end{table*}

\section{Cometary photometry}
In this section we refer to the photometric data base and the criteria applied to construct the cometary lightcurves $m_h$(t). In this regard we analyze some problems in the determination of cometary magnitudes, and discuss how we can relate these magnitudes to the gas production curve.

\subsection{Determination of cometary total magnitudes}

We understand as 'total' magnitudes, the ones that comprise the light coming from the nuclear region and the extended coma. By contrast, the term 'nuclear' magnitude refers to the ones comprising only the central condensation. It should be noted that in most cases the latter magnitudes do not correspond to the bare nucleus, since it is very difficult to avoid contamination by coma light. Nevertheless, efforts are being made to determine true magnitudes of comet nuclei by observing them far from the Sun or by a coma substraction method (e.g. Tancredi {\em et al.} 2006).\\

The precise determination of cometary total magnitudes presents serious problems since comets do not appear as point sources like the stars but as nebulous sources. The measured magnitude will depend on the instrument employed (e.g. Bobrovnikov 1941, Morris 1973). A telescope of small aperture will bring within the field of view the central condensation of the comet plus the extended coma, while a telescope of large aperture will only bring the central condensation. In the latter case the comet will thus look fainter than in the former case. This aperture effect varies both from comet to comet and among different telescopes types. The average aperture effect also depends upon the degree of condensation of the comet. Therefore magnitudes measured by different observers with different instruments can differ by several units of magnitude. As an example, Fig. 1 shows the apparent total magnitudes for the 2005 apparition of comet 9P/Tempel 1, measured by a visual method (i.e. an estimate made with a small telescope or binoculars by eye) as a function of the instrumental aperture. All the magnitude data used in this work comes from the {\em International Comet Quaterly} (ICQ) (courtesy of D. Green), unless we specify another source.  Following the ICQ's recommendations, to minimize the aperture effects the magnitudes should be obtained using the smallest objective aperture and magnification needed to easily see the comet. In Fig. 1 we can see a general trend to reporting brighter magnitudes with smaller apertures, and viceversa. We also note that even for a given aperture, there is a noticeable dispersion in the magnitudes reported by different observers.

We are restricted to the spectral region where the human eye is more sensitive, i.e. all the magnitudes considered in this work are visual magnitudes. So in the following, by 'visual magnitudes' we refer to the cometary magnitudes estimated by a visual method, while by 'CCD magnitudes' we refer to estimates based on a CCD detector with a V broad-band filter or on an unfiltered CCD detector. According to the ICQ, it has been found by some observers that V and unfiltered magnitudes of comets do not differ by more than several tenths of a magnitude. Mikuz and Dintinjana (2001) have compared also CCD V and CCD unfiltered observations finding no meaningful differences among them.

 For faint comets, detectors (e.g. CCDs) tend to record only the nuclear condensation while losing the broad (faint) coma that fades into the sky background. Therefore, the measured 'total' magnitudes will be fainter than they actually are. The question of how CCD magnitudes - whether filtered or unfiltered - relate to visual magnitudes is still open. Towards the limit of visual observations (near magnitudes 12 - 15), it has been noted that CCD total magnitudes are typically 1-3 magnitudes fainter than visual estimates (Green 1996). As an example, Fig. 2a shows the apparent visual magnitudes as compared to the apparent CCD magnitudes, for the 2005 apparition of 9P/Tempel 1. The figure shows that the CCD magnitudes, as a group, are systematically fainter than the visual magnitudes. The CCD magnitudes are also more scattered, while the visual magnitudes depict a more compact group.

To take into account the sky fading effect, Kr\ee sak and Kres\'akov\'a (1989) suggest the following empirical correction for the measured apparent magnitudes of faint comets with $m$ $\ge$ 9

\begin{center}
\begin{equation}
m_{corr} \ = \ 0.5 m + 4.5
\label{kresak}
\end{equation}
\end{center}

\noindent
Figure 2b shows that the corrected CCD magnitudes are much less dispersed  (resulting in a more flat curve). Yet, in spite of a good agreement with respect to the visual magnitudes group, as we go backwards in time ($\gsim$ 50 days before perihelion), we see an increasing departure of the CCD group from the visual group: the CCD lightcurve pre-perihelion branch seems to be significantly different from the visual pre-perihelion branch, with a less steep slope. We conclude that the corrected CCD lightcurve does not agree well with respect to the visual lightcurve.

Since we could not find an adequate correction to CCD magnitudes for the sky fading effect, we use only the visual observations in this work. We also found that for the comets of our sample the visual observations are much more numerous than the CCD observations, during the most active period of the comet (see Fig. 2a for instance).

But the visual observations also show a considerable dispersion, as we saw before (the observations are made with different instrumentation, methodology and/or reference stars, and probably in different local conditions), so we have to adopt some criteria to homogenize the visual magnitude data.

As the first step, we discarded the observations made when the comet was at low altitude without an atmospheric extinction correction applied, or when the magnitude estimate is not very accurate, or when it was made under poor weather conditions according to the information provided by the observer.

 Whereas there are not obvious effects that may lead to an increase in the perceived brightness, other than observation during an outburst, there are several effects (like large apertures or an excess of magnification as we mentioned before, and others like moonlight, twilight, haze, cirrus clouds, dirty optics, etc.) that will decrease the brightness and hence overestimate the observed visual magnitudes. As noted by Ferr\ii n (2005), the top observations follow a smooth trend, defining a sharp boundary, while the lower part of the visual observations is more diffuse, poorly defined (see the visual lightcurve of Tempel 1 in Fig. 2a). Following Rickman {\em et al.} (1987), we take a third, fourth or sixth order polynomial fit to the upper envelope of the ensemble of photometric observations as the lightcurve $m_h(t)$. In practice this polynomial is obtained as a least-square fit to a set of selected data points. This set is defined by the following procedure: we divide the time domain (defined by the observations) in $N$ bins of a fixed size. Then we select, for each one of these time bins, the three brightest observations (namely the ones with the lowest values of the heliocentric magnitude), that do not depart more than 3-sigma from the mean value of the observed magnitudes within the bin. We neglect those time bins with very few observations. Therefore, we allow for a certain range of magnitudes to define our upper envelope, not just the brightest one within each time bin, which gives us more confidence that such a range should cover all potential sources of errors in the magnitude measurements. We used a time bin of 3 days, except for comet 46P/Wirtanen for which we used a time bin of 5 days due to the relatively small number of observations involved.

 The definition of the lightcurve, as a third, fourth or sixth order polynomial, implies that it varies smoothly with time, so comets suspected of having outbursts or another "anomalous" photometric behavior are excluded of our sample. The polynomial fit does not have any symmetry restriction either, so asymmetric fits (with respect to perihelion) are allowed, and actually are usually the case. The only inconvenience we find with this lightcurve definition is that we can not extrapolate the magnitude beyond the time interval defined by the observations, which is equivalent to neglect the gas production outside such an interval. Nevertheless, outside the observational time interval the gas production usually drops several orders of magnitude with respect to the maximum, so we do not expect to introduce a significant bias in the computed mass due to this constraint (in Section 6 we discuss the uncertainty in the computed cometary masses due to the assumptions we make as regards to their lightcurves).

We also neglect short-term variations in the visual magnitudes (like the amplitude of the rotational lightcurve), and variations due to the phase angle, since these sources of intrinsic brightness scattering would be included in the fitted lightcurve by considering sets of the three brightest magnitudes in every time bin, as discussed before.

\begin{figure*}
\centering
\begin{minipage}{140mm}
\includegraphics[width=1.0\textwidth]{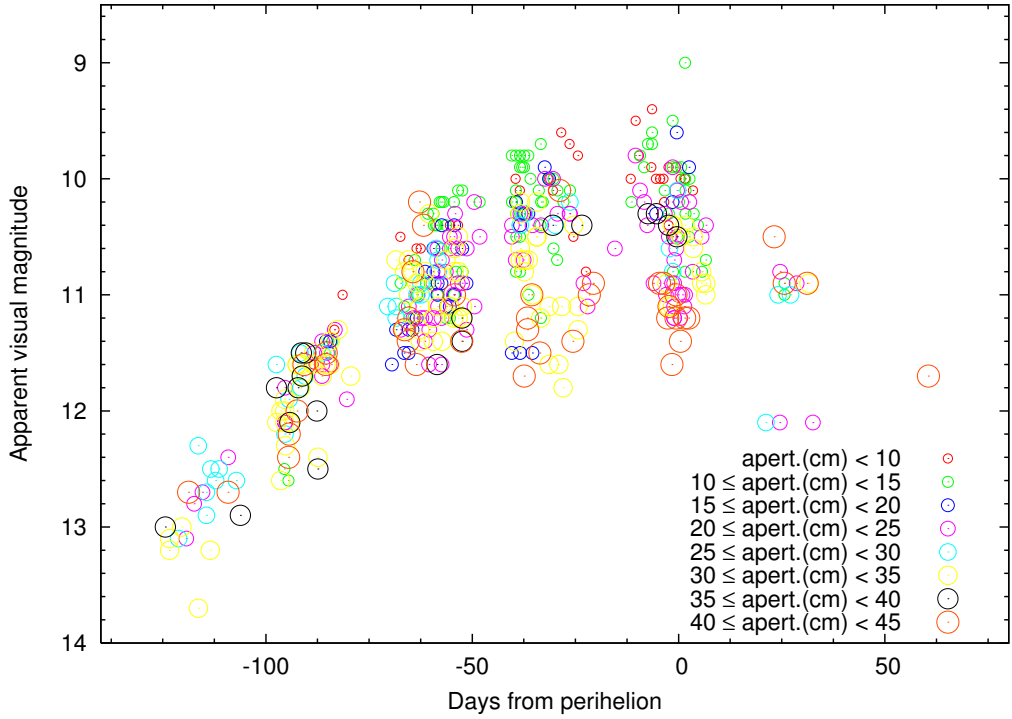}
%\vspace{302pt}
\caption{Apparent visual total magnitudes (circles) as a function of time from the perihelion passage for the 2005 apparition of 9P/Tempel 1 (data from the ICQ archive). The size of the circles indicates the size of the aperture used: the largest circles corresponds to apertures of 40 $\leq$ diameter $<$ 45 cm, while the smallest circles corresponds to apertures of diameter $<$ 10 cm.}
\end{minipage}
\end{figure*}

\begin{figure*}
\centering
\begin{minipage}{140mm}
\includegraphics[width=1.0\textwidth]{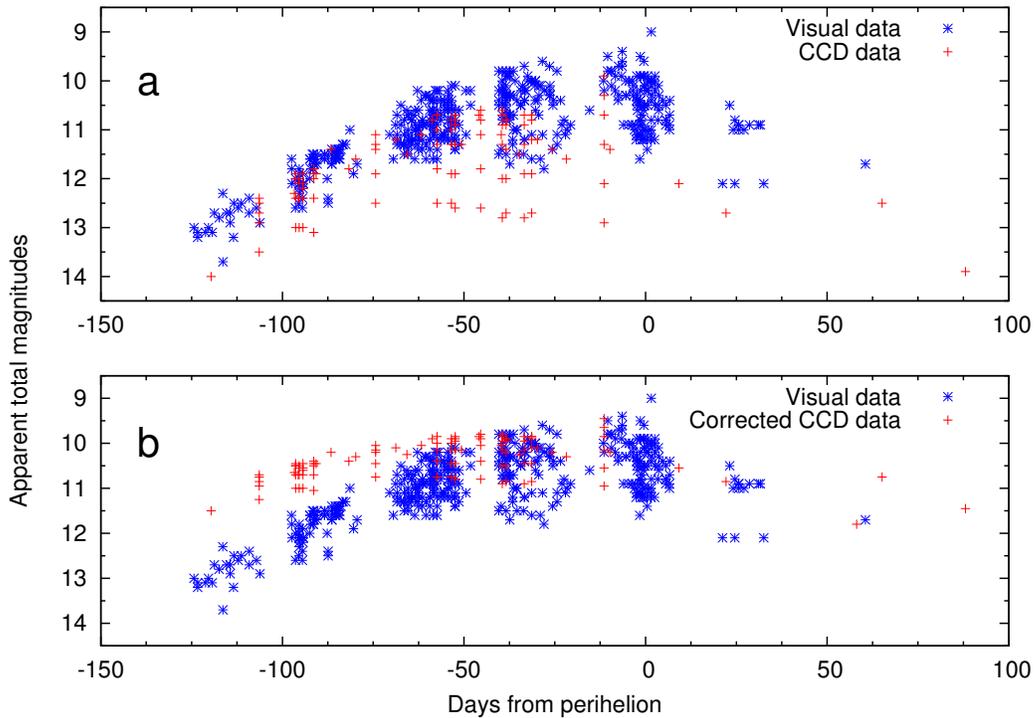}
\caption{Apparent visual total (asterisks) and CCD (plus signs) magnitudes as a function of time (days from the perihelion passage), for the 2005 apparition of 9P/Tempel 1 (data from the ICQ archive).
{\bf (b)}: The same as the previous figure, except that the CCD magnitudes were corrected for the sky fading effect according to Kr\ee sak and Kres\'akova (1989)'s formula.}
\end{minipage}
\end{figure*}

\subsection{Visual magnitudes and gas production rates}

As we pointed out before, at heliocentric distances $r$ $\lsim$ 3 AU the cometary activity is governed by the sublimation of water ice. The gas production rate of water ice is thus an adequate quantitative indicator of cometary activity. Direct observations of the H$_2$O production rate are difficult and sparse because they involved sophisticated and oversubscribed instrumentation. Indirect observations (e.g. observation of water-derived products like the OH radical from its radio lines at 18 cm and from narrow-band photometry in the near UV) are more accessible, but also limited. The huge database of visual magnitudes gives another approach to this problem (Crovisier 2005). In this regard many authors have studied the empirical relation between total visual magnitudes (corrected for geocentric distance) and gas production rates (Festou 1986, Roettger {\em et al.} 1990, Jorda {\em et al.} 1992 and 2008, de Almeida {\em et al.} 1997). From a statistical analysis of 37 comets, based upon $m_h$ values from the ICQ and $Q_{H_2O}$ from the Nan\c{c}ay database, Jorda {\em et al.} (2008) derived the linear relation shown in eq. (\ref{calibracion}) where $Q$ = $Q_{H_2O}$, $a_1$ = -0.2453 and $a_2$ = 30.675. This correlation is based on a larger database than that of the former work of Jorda {\em et al.} (1992), and it would be more reliable (L. Jorda, private communication). This relation (or similar ones) has been used to predict gas production rates of comets in the absence of direct measurements (e.g. Festou {\em et al.} 1990, Jorda and Rickman 1995, de Almeida {\em et al.} 1997). It has also been explored a similar relationship for the CO molecule by Biver (2001), for $r$ $>$ 3 AU.

Festou (1986) argued that the relationship (eq. \ref{calibracion}) seems to hold very well if a set of conditions is fulfilled: 1) a steady state is established in the coma (which would imply that the lightcurve varies smoothly with time), 2) the visible part of the comet spectrum is dominated by C$_2$ emissions, and 3) the ratio of C$_2$ to OH does not vary from comet to comet and with the heliocentric distance. This latter condition could suggest a similar composition for all comets. In this regard A'Hearn {\em et al.} (1995) found that there is little variation of relative abundances with heliocentric distance, and there is also little variation from one apparition to the next for most short-period comets.

In Fig. 3 we show the empirical water production rate $Q(t)$ as a function of time that we obtained for comets 9P/Tempel 1, 19P/Borrelly, 67P/C-G and 81P/Wild 2. Each curve $Q(t)$ was obtained from the respective visual lightcurve (shown in Section 5), where the heliocentric visual magnitudes ($m_h$) were converted to water production rates through eq. (\ref{calibracion}), using Jorda {\em et al.} (2008) calibration's coefficients. Fig. 3 also shows an individual calibration for each comet, obtained from a least-square linear fit between the observed $\log(Q_{H_2O})$  and the estimated magnitudes $m_h$ at the respective observing times (we did not applied any filter or weight to the compiled observational data). We cannot tell at this point if these individual calibrations make a better fit than the Jorda {\em et al.} (2008) calibration, due to the scatter (and relatively scarcity) of the observational measurements.
 We also computed the water production rates predicted by the Jorda {\em et al.} (1992) calibration's coefficients, and found that the residuals between predicted and observed water production rates were about a factor of two larger than the residuals obtained with the Jorda {\em et al.} (2008) constants.
So we chose the Jorda {\em et al.} (2008) calibration as our nominal one, for all comets of our sample, and take the differences in the computed masses (with respect to the computed masses using the individual calibrations), as an estimate of the uncertainty due to the calibration's coefficients (see Section 6.1 for more details on the error's estimate).

\begin{figure*}
\centering
\begin{minipage}{140mm}
\includegraphics[width=1.0\textwidth]{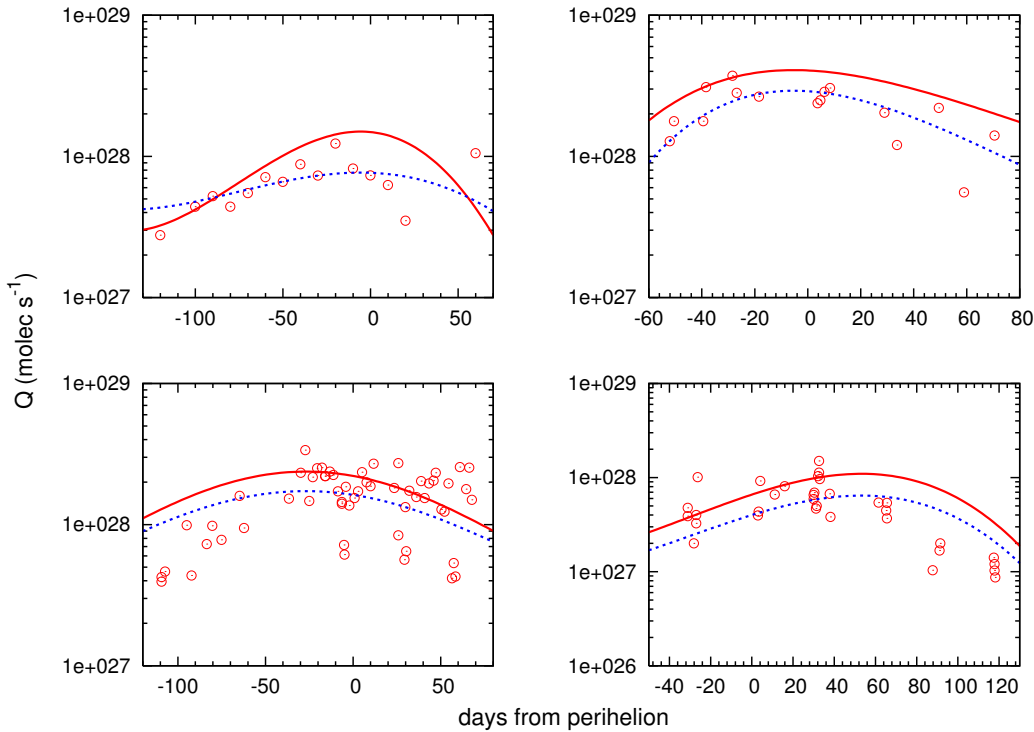}
\caption{Empirical water production rate as a function of time $Q(t)$ (\emph{full line}), superposed to observational data (\emph{small circles}), for the 2005 apparition of 9P/Tempel 1 (\textbf{top, left}), for the 1987, 1994 and 2001 apparitions of 19P/Borrelly (\textbf{top, right}), for the 1982 and 1996 apparitions of 67P/C-G (\textbf{bottom, left}), and for the 1990, 1997 and 2003 apparitions of 81P/Wild 2 (\textbf{bottom, right}). The empirical function $Q(t)$ was obtained from the visual lightcurve (shown in Section 5) via eq. (\ref{calibracion}), using Jorda {\em et al.} (2008) calibration's coefficients. The dotted line shows an individual calibration for each comet, obtained from a least-square linear fit between the observed $\log(Q_{H_2O})$ and the estimated $m_h$ from the lightcurve  at the same observing times. The data corresponds to observed water production rates extracted from Davidsson {\em et al.} (2007) for comet Tempel 1 (courtesy of P. J. Guti\ee rrez), from Davidsson and Guti\ee rrez (2004) for comet Borrelly, from Davidsson and Guti\ee rrez (2005) for comet C-G, and from Davidsson and Guti\ee rrez (2006) for comet Wild 2. }
\end{minipage}
\end{figure*}

We consider, as Crovisier (2005), that the empirical relation (eq. \ref{calibracion}), even if it is physically poorly understood, it is a very useful tool. Maybe the present lack of a physical interpretation of this empirical relationship reflects our poor knowledge of how the cometary activity takes place. By assuming a very simplified physical model for the comet nucleus, that consists of a spherical nucleus with a smooth surface, and that most of the solar radiation is evenly spread across the surface and spent in the sublimation of ices, we then have $Q$ $\propto$ $r^{-2}$ ($r$ $\lsim$ 3 AU), so $\log Q$ $\propto$ -2$\log r$. Since $m_h$ = $H$ + 2.5$n$$\log r$, where $H$ is the absolute total magnitude and $n$ is the photometric index that gives the comet's brightness variation with $r$, we can relate the previous two relations obtaining

\begin{equation}
\log Q  \ \cong \frac{-0.8}{n}\times m_h \ + \ C,
\label{indice}
\end{equation}

\noindent
where $C$ is independent of $r$ and only depends on $H$. Eq. (\ref{indice}) allows us to relate the index $n$ with the coefficient $a_1$ of the empirical calibration (eq. \ref{calibracion}). It is interesting to stress a few consequences of this very simple approach: if the comet brightens exactly like a bare solid body (i.e. the solid nucleus without gaseous activity), we have $n$ = 2, and then $a_1$ = -0.4. But comets usually show activity indices $n$ $>$ 2 which indicates that when approaching the Sun they brighten much more than expected for a bare solid body, because in this case most of the comet light comes from sunlight scattered by dust particles in the coma and fluorescent emission from gaseous molecules, which greatly increases their photometric cross-section. So we expect $n$ $>$ 2 and then $a_1$ $>$ -0.4. This is consistent with the value $a_1$ = -0.2453 from Jorda {\em et al.} (2008), for which we derive $n$ $\sim$ 3. This latter value is closer to the average of the observed slopes of a large number of comets lightcurves ($n$ = 4). As it was pointed out by Biver (2001), the ultimate goal would be to find the scientific justification of this photometric index. This should be related to physical parameters like the amount of dust relative to gas, the size distribution of the dust particles, the relative abundance of C$_2$, etc, as a function of the heliocentric distance. Of course all these implies a much more sophisticated model for the comet nucleus that is beyond the scope of this work.

\section{The studied comets and their lightcurves}

In this section we present the criteria applied to select the comets for this study, and the lightcurves we obtained for the selected comets.\\

We will limit this study to the short-period comets (i.e. $P$ $<$ 200 yr) since these comets have been observed in more than one apparition, so for them it is possible to obtain a good measurement of the non-gravitational effect $\Delta$P\footnote{The evaluation of non-gravitational forces is more difficult for long-period comets, since these have not been observed in a second apparition to check for a change in the orbital period. Nevertheless, non-gravitational terms have been fitted to the equations of motion of several LP comets leading to more satisfactory orbital solutions.}. To select the comets, the following set of conditions was adopted:  1) a measured non-gravitational change $\Delta$P, 2) an adequate photometric coverage during the most active period around the perihelion passage, 3) a good estimate of nucleus size (in order to compute a bulk mass density), and preferentially (but not exclusive) 4) a measured water production rate (in order to check the coefficients for the empirical calibration of eq. \ref{calibracion}). The comets selected were 1P/Halley, 2P/Encke, 6P/d'Arrest, 9P/Tempel 1, 10P/Tempel 2, 19P/Borrelly, 22P/Kopff, 45P/Honda-Mkros-Pajdusakova, 46P/Wirtanen, 67P/Churyumov-Gerasimenko and 81P/Wild 2.

The database of apparent total magnitudes used in this work is from the ICQ archive for observations after 1990, and from the {\em Comet Light Curve Catalogue/Atlas Part. I.} (Kam\ee l 1991a) for observations prior to 1990.

For some comets we use observations involved in more than one apparition to make a composite lightcurve, to compensate for the relative low number of observations in a single apparition. This is possible since no significant variation of the orbital elements is noticed on two or more consecutive observed apparitions of the chosen comets, and also by assuming that the shape of the lightcurve remains almost unchanged during consecutive perihelion passages of the comet.

Figures 4 - 14 show the lightcurves obtained as plots of the total visual heliocentric magnitude $m_h$ as a function of the time relative to the perihelion passage $t$ (the observations were previously filtered according to the criteria explained in Section 4). The horizontal line indicates the {\em cut-off} magnitude $m_{hC}$  (i.e. the polynomial fit is considered a good approximation to the lightcurve only within the interval where $m_h$ $\leq$ $m_{hC}$). For the conversion from apparent to heliocentric magnitudes we compute the comets' distances with the {\em Mercury} orbital integrator (Chambers 1999), and use some routines from Press {et al.} (1992) for function interpolation, and other routines from the NOVAS package of the U.S. Naval Observatory, written by G.H. Kaplan, for the conversion of time from Gregorian to Julian date.

As we discussed in Section 3, most comet lightcurves are asymmetric with respect to the perihelion. As a first approach, the perihelion asymmetries of the comet lightcurves could be mainly related to structural changes of the nucleus that occur during the approach to the Sun, like the expulsion of an insulating dust mantle that would leave exposed areas of fresh ice. This effect, combined with the thermal inertia, would increase the gaseous activity after the perihelion passage. Most of the studied comets effectively show a post-perihelion brightness excess. These are the cases of comets 1P/Halley, 6P/d'Arrest, 10P/Tempel 2, 19P/Borrelly, 45/H-M-P, and 67P/C-G. On the other hand, comets 2P/Encke, 22P/Kopff and 81P/Wild 2 show a pre-perihelion brightness excess. Hence the gaseous activity could have its maximum before perihelion. This could be taken as another evidence of the complexity of the cometary activity, suggesting that other causes, like the concentration of the gaseous sublimation in a few discrete active areas, and the orientation of the spin axis, which may, or may not, favor the illumination of the active areas, may also play a role in the asymmetric behavior.
 Comets 9P/Tempel 1 and 46P/Wirtanen show the most symmetric curve of the sample. Comet Tempel 1 shows a slightly pre-perihelion excess of activity, but in this case the near- and post-perihelion coverage is very poor, due to unfavorable geometrical conditions.

 The maxima of the comet lightcurves seem to occur typically around or about a few days before or after perihelion, although greater departures are common (see for example the lightcurves of comets d'Arrest and C-G), as it was already noticed by Festou (1986).

We note that comet Halley at its maximum brightness is several magnitudes brighter than the rest of the comets of the sample at their maxima. This should be correlated with a much greater gaseous activity for Halley as compared to the other comets. For instance, Halley shows a gas production rate of about 10$^{31}$ s$^{-1}$ near perihelion, while the rest of the studied comets are approximately in the range [0.1 - 2.0]$\times$10$^{28}$ s$^{-1}$ near perihelion (A'Hearn {\em et al.} 1995, Crovisier {\em et al.} 2002). This result may be partly due to the different comet sizes, and partly due to the different source regions: while Halley is the prototype of the population of the {\em Halley-Type} (HT) comets - for which the source region would be the Oort Cloud (Fern\'andez 2005, Ch. 7), the rest of the studied comets belongs to the {\em Jupiter Family} (JF) population - for which the main source region would be the trans-neptunian belt (Fern\'andez 1980, Duncan {\em et al.} 1988). Since JF comets may have spent up to hundreds or thousands of revolutions on short-period orbits near the Sun, they may be more physically evolved than HT comets, so 1P/Halley may still be a comet with a fresher and more active surface, leading to its observed substantially higher activity.

Because of their short orbital periods ($P$ $<$ 20 yr), most of the discovered JF comets have been observed in several apparitions, and in some cases through all the orbit up to their aphelia, allowing to gather a very valuable wealth of physical data. Furthermore, several JF comets have been or will be the targets of space missions, which will help to greatly increase our knowledge about their physical nature. This is the case of four out of the eleven comets studied in this work. In the following we make a brief summary for each one of the lightcurves obtained.

\subsection{1P/Halley}

Comet 1P/Halley was the target of several international space missions during its last apparition in 1986, which could take images of its nucleus of irregular shape, and confirmed, in general terms, Whipple's model. From the ground the comet was bright enough to be visible to the naked eye, though far below the spectacular previous apparition of 1910. This comet presents the best photometric coverage in one single apparition, as compared to the rest of the studied comets, resulting in a good quality lightcurve. The plot in Fig. 4 contains 1221 observations corresponding to the 1986 apparition, and shows the fitted lightcurve $m_h(t)$. The lightcurve obtained is similar to others found in the literature (e.g. Ferr\ii n 2005).

\subsection{2P/Encke}

For this comet we combined the observations of four consecutive apparitions: 1990 (186 observations), 1994 (202 observations), 1997 (29 observations), 2000 (24 observations). The observations of 1997, in spite of their low number, were important to define better the post-perihelion branch, for which only the 1994 apparition has post-perihelion observations. Figure 5 shows a good consistency between the different apparitions, confirming our assumption that the lightcurve has not changed significantly from one apparition to the next, during the last decade. Yet we observe a secular decrease of the non-gravitational effect during this period. This was already noticed by Kam\'el (1991b) who studied the lightcurve evolution of this comet from 1832 to 1987. Kam\'el also found a secular decrease of the perihelion asymmetry (which he related to a shift in the time of maximum brightness from 3 weeks before perihelion to a few days after perihelion in the latest apparitions). He concluded that the shape of the lightcurve has not significantly changed during short intervals of time (like the last apparitions he studied), in particular the perihelion asymmetry has remained, showing a faster brightness decrease after the maximum and a slower increase before it. Our composite lightcurve, based on most recent observational data, is in good agreement with the results from Kam\'el.

\subsection{6P/d'Arrest}

The lightcurve dataset for this comet contains 424 observations from the 1995 apparition (the last apparition in 2002 was unfavorable since the comet was in conjunction with the Sun close to the time of the perihelion passage).
The lightcurve is shown in Fig. 6. This comet shows an extraordinary strong asymmetry with respect to perihelion. The brightness rises rapidly as the comet passes perihelion, reaching the maximum around $\sim$ 40 days after perihelion, with a slow decline after the maximum. The lightcurve obtained is similar to others found in the literature (e.g. Szutowicz and Rickman 2006).

\subsection{9P/Tempel 1}

This comet has been the most recent target of a successful space mission, the {\em Deep Impact}, which encountered the comet about 1 day before perihelion on July 4,  2005. Therefore a great amount of data has been collected for this comet, including a good photometric coverage. Unfortunately, like the former apparition of 1994, there are very few post-perihelion observations due to geometrical circumstances. The lightcurve dataset for this comet contains 447 observations from the 2005 apparition. The lightcurve is shown in Fig. 7. We also studied the lightcurve corresponding to the 1994 apparition (with a dataset of 603 observations), and found an average brightness decrease of about 1-2 mag from 1994 to 2005. The perihelion asymmetry seems to remain, with a slightly pre-perihelion predominance, but this feature is quite uncertain since there are not enough post-perihelion observations to define a good post-perihelion lightcurve. Ferr\ii n (2005, 2007) has published a lightcurve for this comet, but only based on the 1994 data, which is similar to the 1994 lightcurve we obtained.

\subsection{10P/Tempel 2}

The lightcurve dataset for this comet contains 153 observations from the 1983 apparition and 143 observations from the 1988 apparition. The composite lightcurve is shown in Fig. 8. There are few pre-perihelion observations in 1983, but those from 1988 are enough to properly define the pre-perihelion branch of the lightcurve. We found a good consistency between the observations of both apparitions, which implies that the lightcurve remains practically the same. Rickman {\em et al.} (1991a) have studied the lightcurve of this comet during 13 consecutive apparitions from 1899 up to 1988. They find that after perihelion the 1983 and 1988 apparitions yield quite different lightcurves. The disagreement between their result and ours may be due to a different filtering criteria applied to the observations (since the original database - Kam\'el 1991a - is the same). In correspondence with a negligible variation of the lightcurve between both apparitions, we do not find any significant change in the orbital parameters nor in the non-gravitational effect (see Tables 2 and 3, respectively). Anyway, both works agree in the remarkable asymmetry showed by this comet lightcurve, which is very similar to the d'Arrest case: a rapid brightness increase before the maximum, followed by a slow decrease after it.

We also studied the 1999 lightcurve, based on 239 observations (the 1994 and 2005 apparitions were too poor in number of observations). We found a significant brightness decrease of 1-2 mag from 1988 to 1999, and a shift in the time of the maximum brightness from around 14 days after perihelion to a few days after perihelion. Notwithstanding this remarkable change in the lightcurve, the asymmetry remains in shape: faster brightness increase before the maximum, and a slower brightness decrease after it (although the pre-perihelion increase is less steep than that of the 1988 lightcurve). This brightness decrease could be related to a small but noticeable increase (of about 7 \%) of the perihelion distance from 1988 to 1999 (see Table 2). We also noticed a certain decrease (of about 20 \%) of the non-gravitational change $\Delta$P (see Table 3). This comet shows the smallest non-gravitational effect of the sample (closely followed by Tempel 1). Rickman {\em et al.} (1991a) suggested a possible correlation between  the lower values of the non-gravitational effect on Tempel 2 and the apparent drop in the gas production as inferred from the lightcurves.

We conclude that there is a remarkable difference of the 1983 and 1988 composite lightcurve with respect to the 1999 lightcurve. Since the quality of the former is much better, we adopt it as our nominal lightcurve (a short explanation of how we evaluate the lightcurve quality is given at the end of this section).

\subsection{19P/Borrelly}

This comet has been the target of the {\em Deep Space 1} mission, which encountered the comet on September 22, 2001, about 8 days after perihelion.
The lightcurve dataset for this comet contains 360 observations from the 1987 apparition, 853 observations from the 1994 apparition, and 242 observations from the 2001 apparition. The composite lightcurve is shown in Fig. 9. We found no meaningful changes in the lightcurve from one apparition to the next, during the studied period. We discarded a small group of bright measurements of the 2001 apparition bunched together around 20 days before perihelion as a sort of spike, since they depart from the general trend of the bulk of the observations. This spike could be due to a little outburst produced around 30 to 20 days before perihelion. Borrelly shows the same asymmetrical behavior as, for example, d'Arrest or Tempel 2: a faster brightness increase before the maximum followed by a slower decrease after it. The composite lightcurve obtained in this work resembles very much the one from Ferr\ii n (2005).

\subsection{22P/Kopff}

The lightcurve dataset for this comet contains 349 observations from the 1983 apparition and 586 observations from the 1996 apparition. The composite lightcurve is shown in Fig. 10. This comet presents a curious lightcurve since, even though it shows a faster brightness increase before the maximum, like those comets with a post-perihelion excess (e.g. C-G, d'Arrest, Tempel 2 or Borrelly), it shows however a pre-perihelion excess. We note however that the brightness maximum is quite uncertain, since the comet seems to hold a similar activity level for several weeks, so we cannot exclude the possibility that the real maximum  could be a few days after perihelion. Anyway, it is clear from Fig. 10 that the comet shows more activity before than after perihelion.

\subsection{45P/Honda-Mkros-Pajdusakova}

The lightcurve dataset for this comet contains 117 observations from the 1990 apparition, 108 observations from the 1995 apparition, and 40 observations from the 2001 apparition. The composite lightcurve is shown in Fig. 11. This comet shows a post-perihelion excess but not so strong as C-G, d'Arrest, Tempel 2 or Borrelly. We can see an increasing scatter of the observations as the comet moves away from perihelion (the post-perihelion observations after about 60 days were not considered for this reason).

\subsection{46P/Wirtanen}

This comet was the former target of the {\em Rosetta} mission, which will finally be encountering comet C-G in 2014. The lightcurve dataset for this comet contains 105 observations from the 1991 apparition. The lightcurve is shown in Fig. 12. We also studied the lightcurve corresponding to the 1997 apparition (with a dataset of 104 observations). We note that the 1997 observations seems to be somewhat brighter (about several tenths of magnitude) than those from 1991, so we decided not to combine both apparitions into one single composite lightcurve, since it is not clear that the lightcurve has not changed from one to the other. Since the 1997 observations are much more scattered than those from 1991 as we go far from the perihelion, which results in an more poorly defined lightcurve, we decided to take the 1991 lightcurve as the nominal one.  As inferred from this lightcurve, besides a post-perihelion brightness excess, this comet also shows a rapid decrease of the gaseous activity for increasing heliocentric distances. This is in agreement with Jorda and Rickman (1995), who also studied the 1991 lightcurve of this comet.

\subsection{67P/Churyumov-Gerasimenko}

This comet will be encountered by the {\em Rosetta} spacecraft in 2014, which will orbit it during its journey through the inner solar system, and land a probe on its surface. Hence a great wealth of physical data for this particular comet is expected in the next years.

The lightcurve dataset for this comet contains 260 observations from the 1982 apparition, and 127 observations from the 1996 apparition. The composite lightcurve is shown in Fig. 13. This comet shows a strong post-perihelion asymmetry. We also note that this comet is the faintest one of the sample (followed closely by Tempel 1), since its maximum brightness is somewhat below 10 mag, while most JF comets show maxima at magnitudes between 7 - 9, as we pointed out before. The post-perihelion excess and the maximum brightness obtained is in agreement with other works (e.g Ferr\ii n 2005).

\subsection{81P/Wild 2}

This comet was visited by the {\em Stardust} spacecraft in January 2, 2004, which obtained high resolution images of the nucleus, showing remarkable differences with previous images of Halley and Borrelly, like the evidence of impact craters (Brownlee {\em et al.} 2004).

The lightcurve dataset for this comet contains 78 observations from the 1990 apparition, 626 observations from the 1997 apparition, and 72 observations from the 2003 apparition. The composite lightcurve is shown in Fig. 14. This comet shows the most strong pre-perihelion asymmetry of the sample. The bulk of the observations which defines the lightcurve around the maximum brightness corresponds to the 1997 apparition, while the 2003 observations contributes to define the pre-perihelion branch, and the 1990 observations help to define the outermost part of the post-perihelion branch. This is the only comet of the sample which has a visual photometric coverage up to heliocentric distances of about 3 AU. Our lightcurve is in good agreement with that studied by Sekanina (2003). Ferr\ii n (2005) also derived a similar lightcurve for this comet, except that ours shows a steeper post-perihelion slope, which implies a more rapid fading of the activity.

\begin{figure*}
\centering
\begin{minipage}{140mm}
\includegraphics[width=1.0\textwidth]{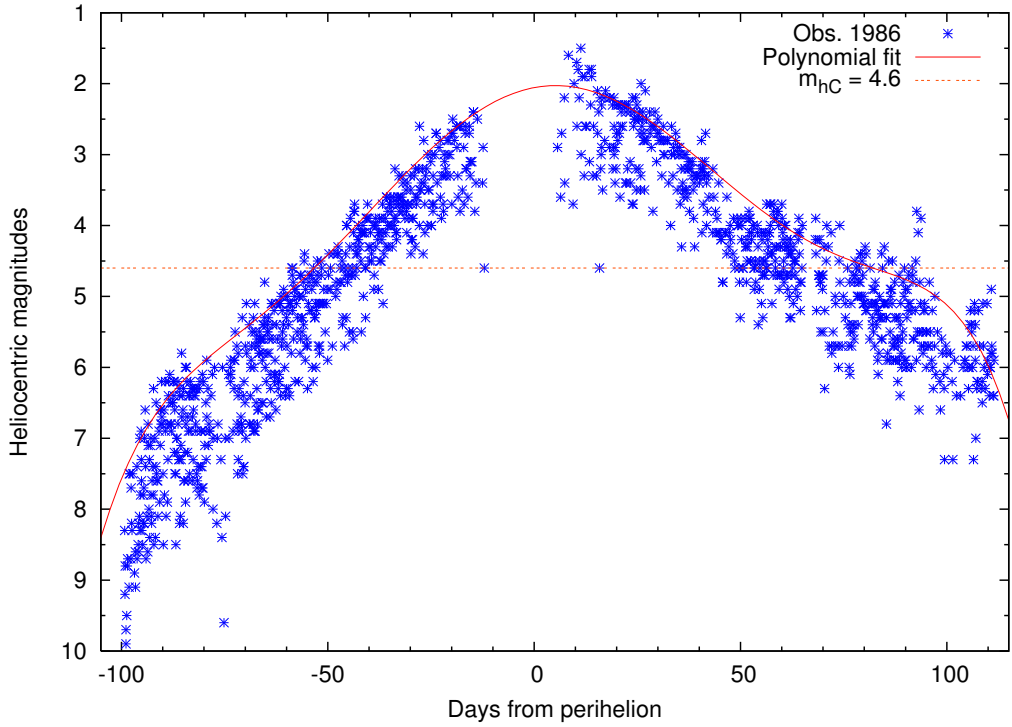}
\caption{Heliocentric total visual magnitudes as a function of time (days from the perihelion passage), for comet 1P/Halley. The polynomial fit $m_h(t)$ to the upper envelope of the broad distribution of photometric measurements is shown. The horizontal line indicate the {\em cut-off} magnitude $m_{hC}$, as defined in Section 5. The value of $m_{hC}$ is also shown. The magnitude data for this comet and the rest of the sample were extracted from Kam\'el 1991a (for observations before 1990) and from the ICQ archive (for observations from 1990 and forward).}
\end{minipage}
\end{figure*}

\begin{figure*}
\centering
\begin{minipage}{140mm}
\includegraphics[width=1.0\textwidth]{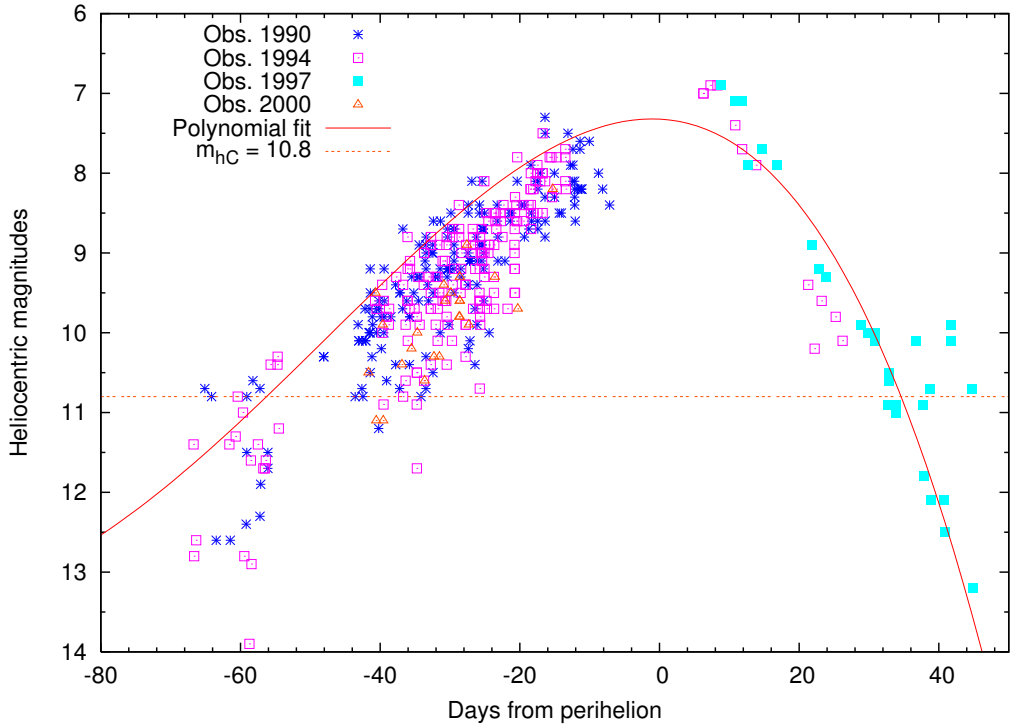}
\caption{Idem as Fig. 4 for comet 2P/Encke.}
\end{minipage}
\end{figure*}

\begin{figure*}
\centering
\begin{minipage}{140mm}
\includegraphics[width=1.0\textwidth]{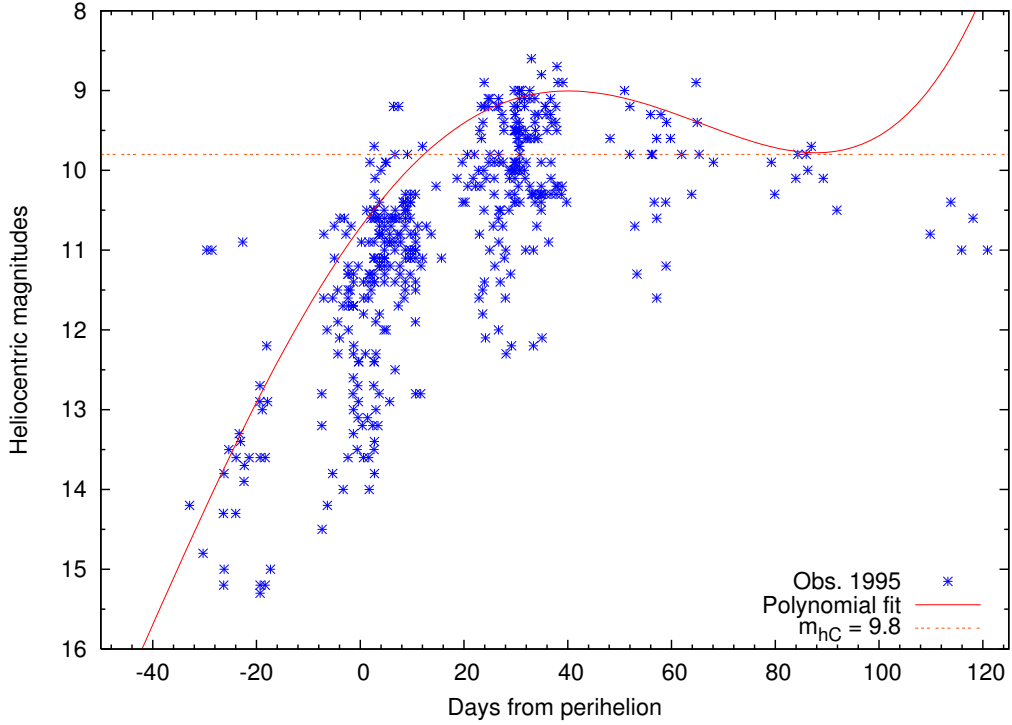}
\caption{Idem as Fig. 4 for comet 6P/d'Arrest.}
\end{minipage}
\end{figure*}

\begin{figure*}
\centering
\begin{minipage}{140mm}
\includegraphics[width=1.0\textwidth]{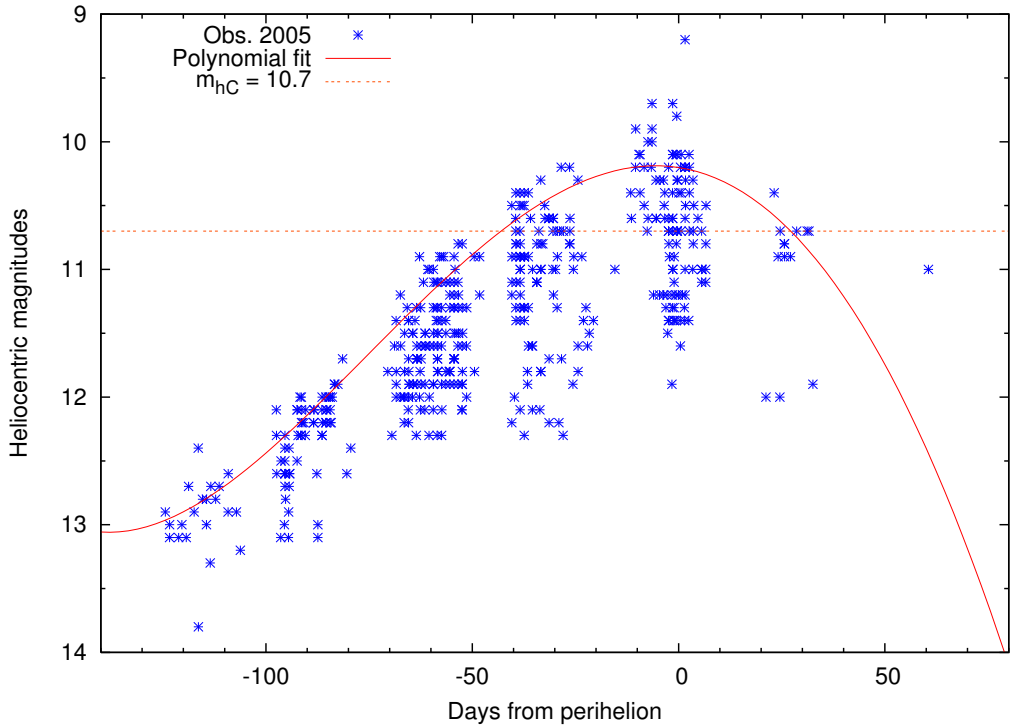}
\caption{Idem as Fig. 4 for comet 9P/Tempel 1.}
\end{minipage}
\end{figure*}

\begin{figure*}
\centering
\begin{minipage}{140mm}
\includegraphics[width=1.0\textwidth]{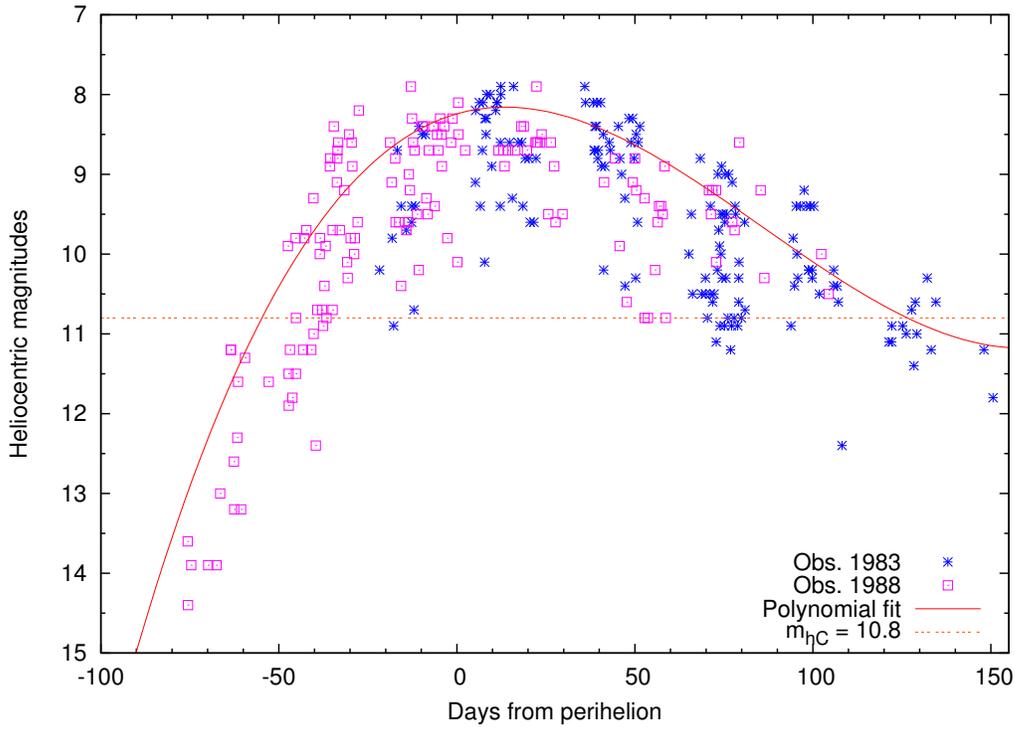}
\caption{Idem as Fig. 4 for comet 10P/Tempel 2.}
\end{minipage}
\end{figure*}

\begin{figure*}
\centering
\begin{minipage}{140mm}
\includegraphics[width=1.0\textwidth]{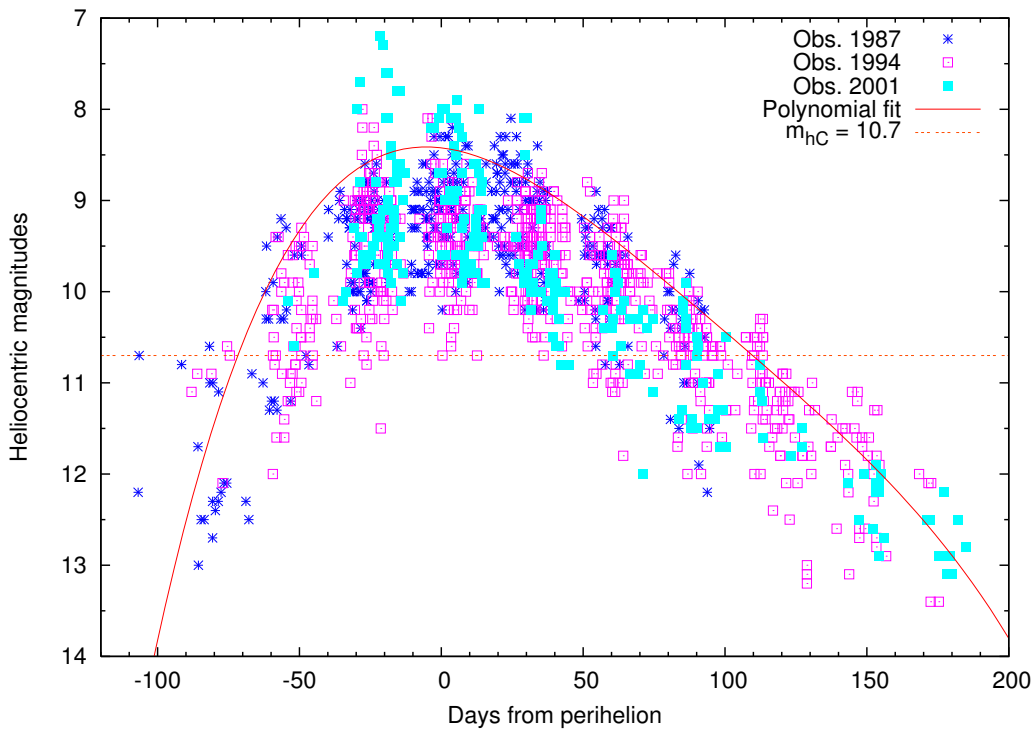}
\caption{Idem as Fig. 4 for comet 19P/Borrelly.}
\end{minipage}
\end{figure*}

\begin{figure*}
\centering
\begin{minipage}{140mm}
\includegraphics[width=1.0\textwidth]{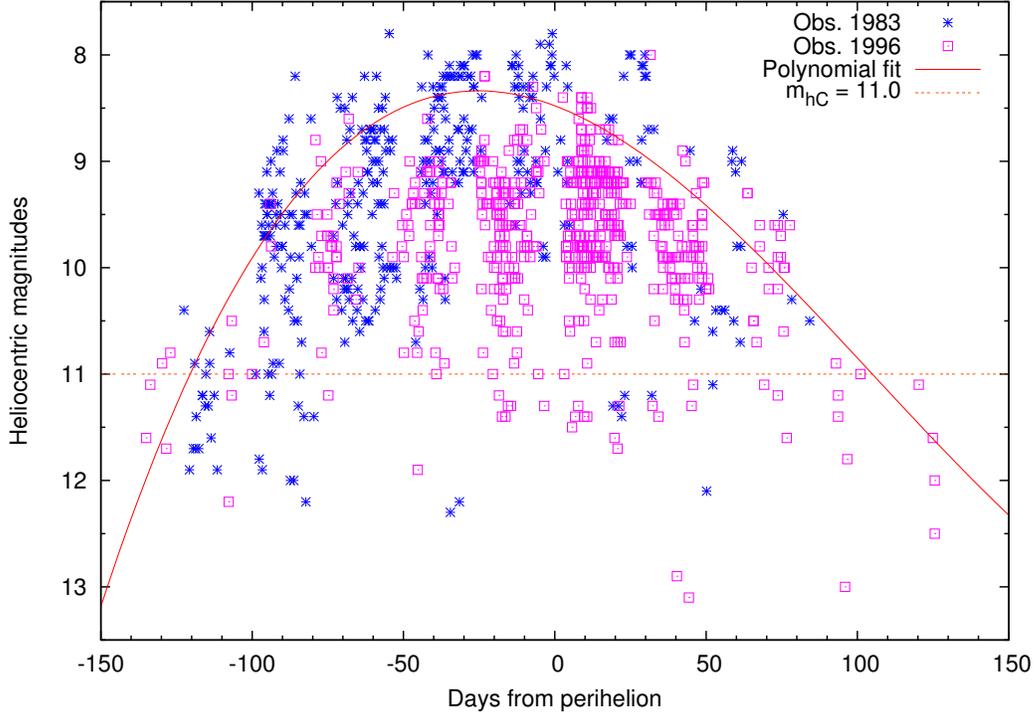}
\caption{Idem as Fig. 4 for comet 22P/Kopff.}
\end{minipage}
\end{figure*}

\begin{figure*}
\centering
\begin{minipage}{140mm}
\includegraphics[width=1.0\textwidth]{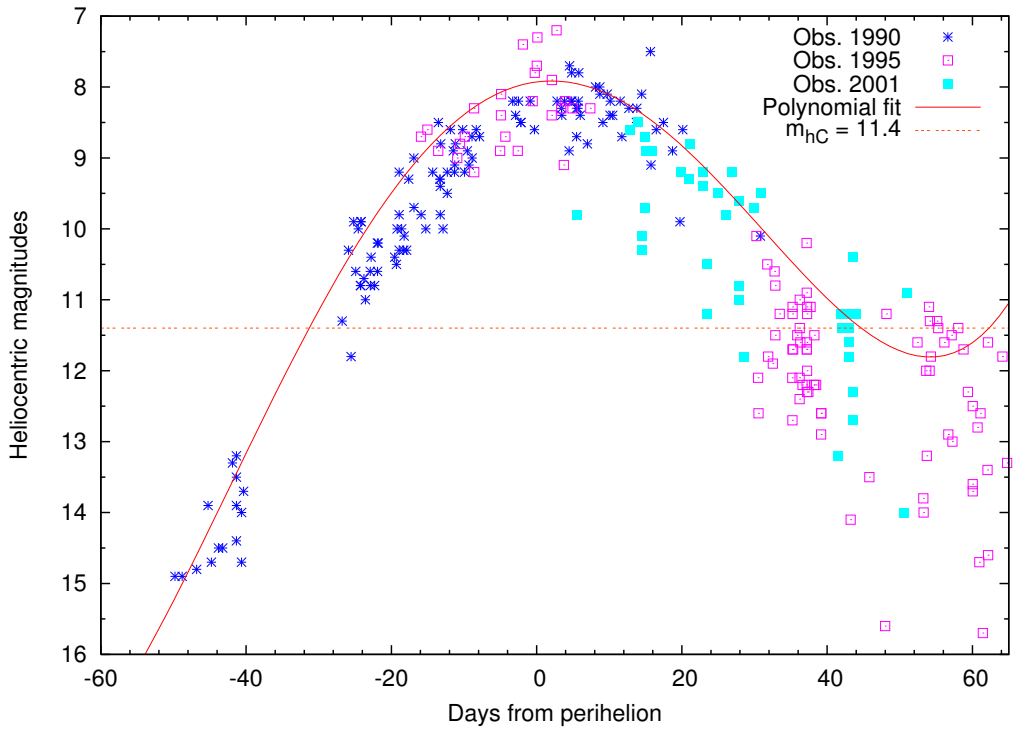}
\caption{Idem as Fig. 4 for comet 45P/H-M-P.}
\end{minipage}
\end{figure*}

\begin{figure*}
\centering
\begin{minipage}{140mm}
\includegraphics[width=1.0\textwidth]{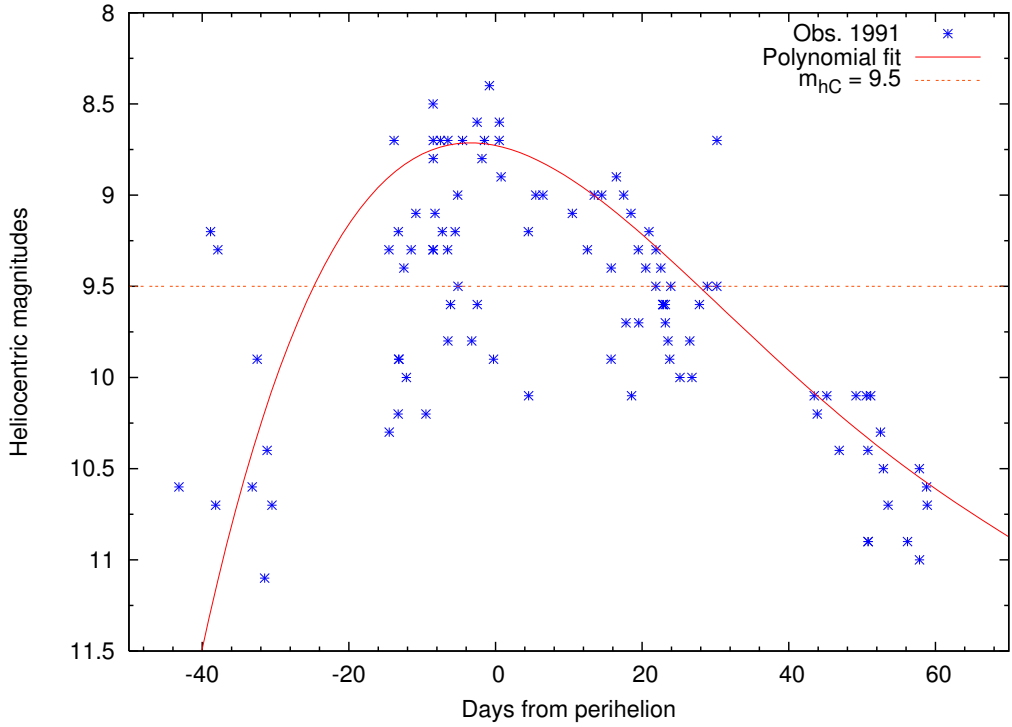}
\caption{Idem as Fig. 4 for comet 46P/Wirtanen.}
\end{minipage}
\end{figure*}

\begin{figure*}
\centering
\begin{minipage}{140mm}
\includegraphics[width=1.0\textwidth]{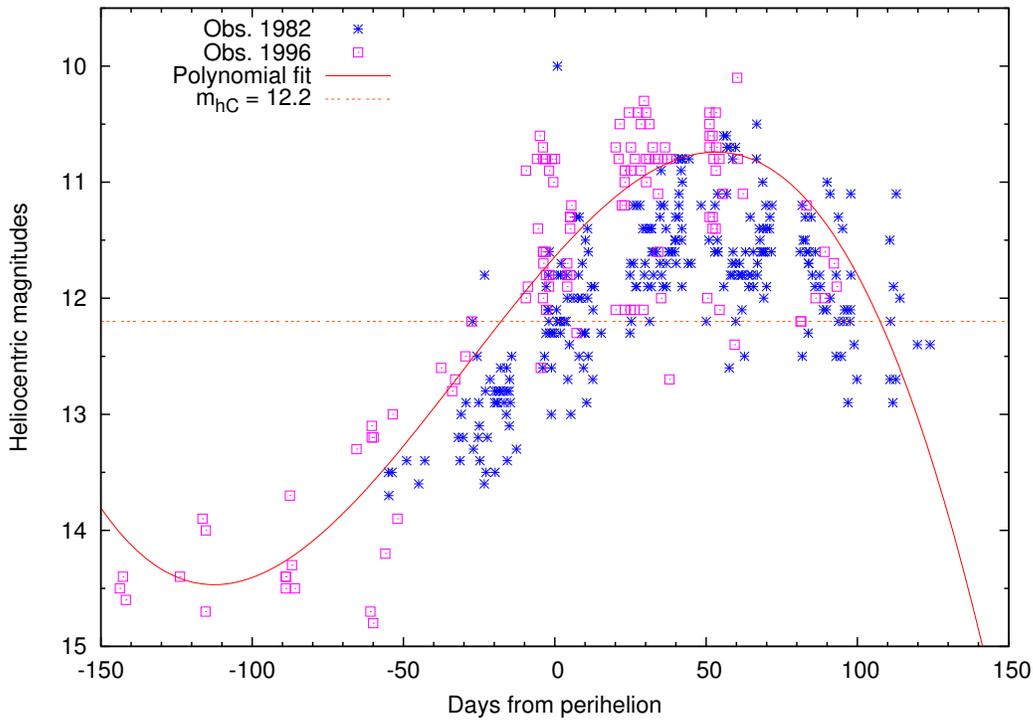}
\caption{Idem as Fig. 4 for comet 67P/C-G.}
\end{minipage}
\end{figure*}

\begin{figure*}
\centering
\begin{minipage}{140mm}
\includegraphics[width=1.0\textwidth]{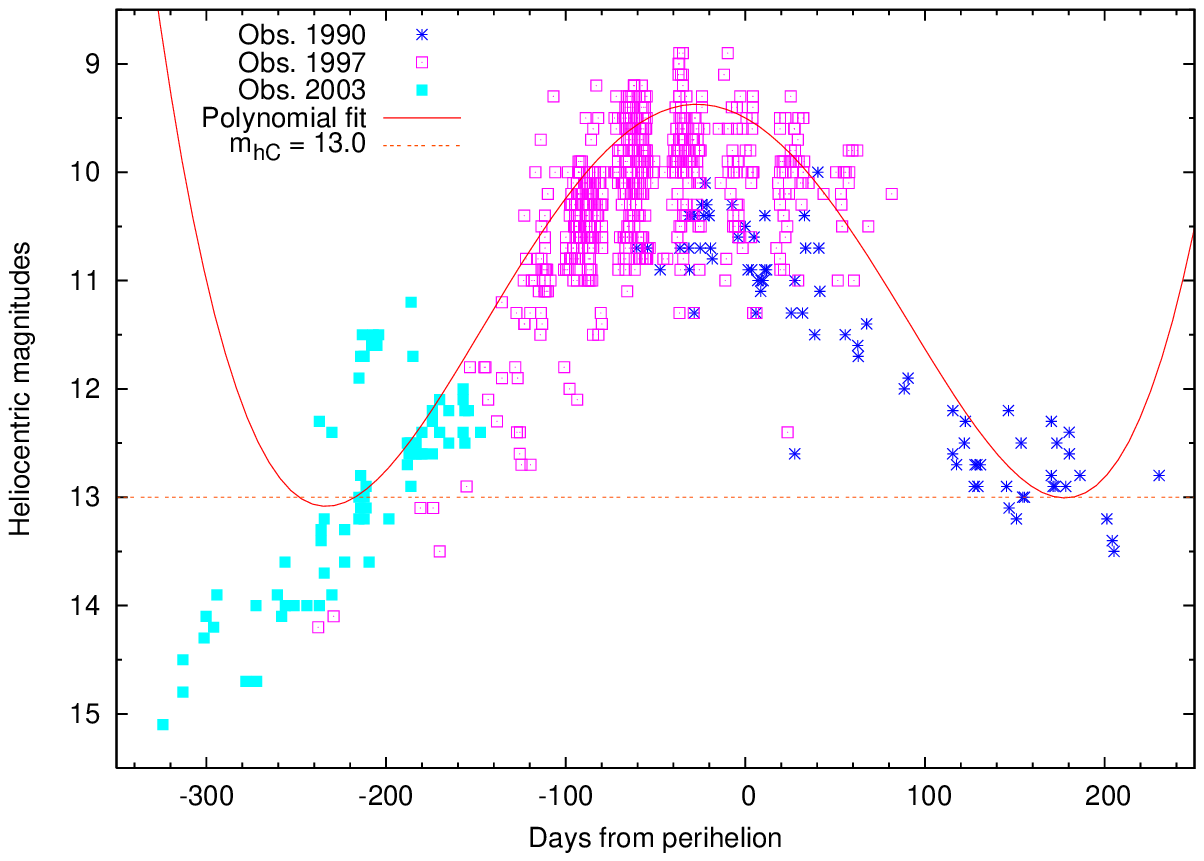}
\caption{Idem as Fig. 4 for comet 81P/Wild 2.}
\end{minipage}
\end{figure*}

\section{The results}

For the computation of the non-gravitational change $\Delta$P and the comet mass (from eqs. (\ref{masa-aprox}) and (\ref{marsden}), respectively) we use some routines from Press {\em et al.} (1992) for numerical integration of functions and for root finding, and write our own routine to solve the Kepler equation.

We first introduce the main sources of uncertainty in the computation of the cometary masses, and then the results obtained for the comets of the sample. Our intention was not to proceed to a rigorous computation of errors, since that would be meaningless under the general assumptions we made and the statistical scope of this work, but instead to evaluate in an empirical way the effect of the main sources of uncertainty in the computed masses.

\subsection{Sources of uncertainty}

The most important sources of uncertainty are the comet lightcurve, the model parameters (particularly the effective outgassing velocity, which maybe is the most uncertain of the parameters assumed), and the computed non-gravitational change $\Delta$P.

\subsubsection{The lightcurve}

In order to estimate the uncertainty due to the lightcurve, we proceeded as follows: as we explained in Section 4.1, the lightcurve is defined by a least-squared fitted polynomial to a set of selected observations. This set of observations was obtained by dividing the observational time interval in bins of fixed size, and selecting, in each bin, the three brightest observations with the condition that they do not significantly depart from the bulk of observations. By varying the bin size (from 2 to 6 days, with one day step), we picked different sets of brightest magnitudes leading to different lightcurves (one per bin size), and hence different \emph{\textit{cut-off}} magnitudes ($m_{hC}$). For each one of these lightcurves (with its corresponding \emph{cut-off} magnitude) we computed the mass (leaving the rest of the parameters involved fixed with their nominal values). In this way we obtained a set of different mass estimates for each comet of the sample. Hence, for a given comet, we take the mean value of the relative differences in the computed masses, with respect to the nominal value, as an empirical estimate of the uncertainty in the comet mass due to the lightcurve uncertainty.

Comet d'Arrest was found to be the comet with the smaller uncertainty due to its lightcurve (($\Delta M/M)_{lightcurve}$ = 3 \%), followed by, in increasing order of uncertainty: Halley and C-G (6 \%), Tempel 1 and Tempel 2 (8 \%), Kopff (9\%), Encke (11 \%), Borrelly (13 \%), Wild 2 (28 \%), H-M-P (43 \%) and Wirtanen (44 \%).

We note that these values do not necessarily reflect the quality of the lightcurve in all cases, since, for instance, Tempel 1 has a relatively poorly defined post-perihelion branch of the lightcurve, though we got a small uncertainty in the computed mass. Nevertheless, in most cases, we find a good correlation between the lightcurve quality and the estimated uncertainty in the mass, as in the case of Wirtanen. This comet shows the most poorly defined lightcurve of  the sample, due to a relatively low number of observations, which reflects in the largest estimated uncertainty in the mass.

\subsubsection{The model parameters}

From the discussion of Section 3, we assumed an error of $\pm$ 0.1 km s$^{-1}$ in $<u>$. We also assumed 0.05 $\leq$ $<|\sin(\eta)\cos(\phi)|>$ $\leq$ 0.15, and 0.9 $\leq$ $<\cos(\eta)>$ $\leq$ 1.\\

Although a thorough study of the correlation between water production rates and visual magnitudes is beyond the scope of this work, we perform individual calibrations for some comets of our sample (as explained in Section 4.2), just to evaluate, in an empirical way, how the uncertainty related to the constants \{$a_1$, $a_2$\} propagates to the computed masses. We found that the relative difference in the mass computed using the individual calibration, with respect to the nominal mass (i.e. the mass computed using the Jorda {\em et al.} (2008) calibration) varied between 13 \% (for Wild 2) and 46 \% (for Tempel 1). The average of these relative differences was about 30 \%.
 We also found that the relative difference in the mass computed using the Jorda {\em et al.} (1992) calibration, with respect to the nominal mass, was about 30 \% for the comets of the sample, except for comet Halley for which the difference was about 20 \%.
 Hence, we estimate the uncertainty in the computed masses due to the calibration constants \{$a_1$, $a_2$\} to be about ($\Delta M/M)_{calibration}$ $\cong$ 30 \%, for all the comets of the sample.\\

\subsubsection{The non-gravitational effect}

For some comets we found in the literature an estimate of the absolute error $\Delta(\Delta P)$ (e.g. Borrelly by Davidsson and Guti\ee rrez 2004, Tempel 1 by Davidsson {\em et al.} 2007, C-G by Davidsson and Guti\ee rrez 2005, and Wild 2 by Davidsson and Guti\ee rrez 2006). For the rest of the comets we have to adopt some criteria in order to estimate the uncertainty of this parameter. For those comets which have showed some variation in the non-gravitational effect $\Delta$P during the apparitions considered in this work (e.g. Encke, Kopff, H-M-P, Wirtanen), we estimate the error as the standard deviation with respect to the mean value (although for these comets intrinsic variations of the change $\Delta$P cannot be ruled out, we neglect them by considering a composite lightcurve). For comets Halley and d'Arrest, in the absence of a better criterion, we assume an error given by the average of the normalized (i.e. relative) errors estimated for the former comets. For comet Tempel 2, which shows a very small $\Delta$P like Tempel 1, we assumed the same error than for this latter comet. The estimated uncertainties range from  $\sim$ 4 \% for comets Kopff and H-M-P, up to $\sim$ 58 \% for comet Wild 2. The intermediate values of the estimated uncertainty are: $\sim$ 6 \% (comets Halley, d'Arrest and Borrelly), $\sim$ 13 \% (comet Wirtanen), $\sim$ 20 \% (comet C-G), $\sim$ 27 \% (comets Tempel 1 and Tempel 2), and $\sim$ 50 \% (comet Encke).

\subsubsection{Estimate of the overall error in the mass and density}

Although for some parameters involved in the computation of the mass (like the non-gravitational effect $\Delta$P, and the effective outflow velocity $<u>$) their effect in the mass can be easily evaluated separately, since the mass is proportional to them, we found more convenient to evaluate the overall uncertainty in the computed mass in an empirical way. In this regard we considered, for each comet of the sample, $N$ points of a parametric space defined by the parameters $<u>$, $<\cos(\eta)>$, $<|\sin(\eta)\cos(\phi)|>$, $\Delta P$, and $F$, where $F$ is a factor that accounts for the uncertainty due to the lightcurve and the calibration constants \{$a_1$, $a_2$\}, defined as $F  =  1  +  (\Delta M/M)_{lightcurve}  +  (\Delta M/M)_{calibration}$ (a value of $F$ = 1 implies a negligible uncertainty due to the lightcurve and the calibration constants). We chose $N$ = 1000. We denote by $F_L$ the computed values of $F$ for the comets of the sample. We obtained values ranging between $F_L$ = 1.3 for comet d'Arrest and $F_L$ = 1.7 for comets H-M-P and Wirtanen. Each $point$ of the parametric space was obtained by generating a random value with an uniform distribution for each parameter, with the constrains: 0.17 $\leq$ $<u>$  $\leq$ 0.37, 0.9 $\leq$ $<\cos(\eta)>$ $\leq$ 1, 0.05 $\leq$ $<|\sin(\eta)\cos(\phi)|>$  $\leq$ 0.15, $\Delta P_{min}$ $\leq$ $\Delta P$ $\leq$ $\Delta P_{max}$, and 1 $\leq$ $F$ $\leq$ $F_L$, where $\Delta P_{min}$ = $\Delta P_{nom}$-$\Delta(\Delta P)$, and $\Delta P_{max}$ = $\Delta P_{nom}$+$\Delta(\Delta P)$, according to the estimated absolute error $\Delta(\Delta P)$ (computed from the relative errors presented in Section 6.1.3), and to the nominal value $\Delta P_{nom}$ for the non-gravitational effect given in Table 3.
Then, each one of these $N$ points represents a set of parameter values from which a mass is computed, by means of eq. (\ref{masa-aprox}), where the factor $F$ multiplies the right hand member of this equation. By computing the standard deviation of these $N$ mass estimates with respect to the nominal value we evaluate the overall uncertainty or error in the mass ($\Delta M$) of a given comet.

Finally, by propagating the estimated errors in the mass ($\Delta M$) and in the effective nucleus radius ($\Delta R$) (given in Sections 6.2.1 - 6.2.11), we obtain the error in the mass density

 \begin{equation}
 \Delta \rho \ = \ \frac{3}{4 \pi} \Big[\frac{\Delta M}{R^3} + \frac{M \Delta R}{R^4}\Big],
 \end{equation}

 \noindent where $R$ represents the effective nuclear radius\footnote{The effective nuclear radius defines the radius of a sphere whose volume is equal to that of the comet nucleus (usually modeled as a triaxial ellipsoid of semiaxes a, b, c)}.

\subsection{Computed masses and densities}

We present the computed masses from the nominal values for the model parameters (summarized in Table 1), with a rough estimate of the mass uncertainty due to the main sources of uncertainty as considered before. We also show the derived bulk mass density with an error estimate as derived from the size and the mass uncertainties, respectively. The results are compared with those from other works, and summarized in Table 5.

\subsubsection{1P/Halley}

{\em Mass.} We derive a mass of [3.2 $\pm$ 1.2] $\times$ 10$^{14}$ kg for this comet.

\noindent
{\em Size and density.} This comet has a determination of its nucleus size measured on close-up images from fly-by missions, which is the best procedure to determine unambiguously the size, albedo, or shape of a comet nucleus. Keller {\em et al.} (1987) determined an effective radius of 5.2 $\pm$ 1.0 km. According to this value for the comet size we derive a bulk density of 0.5 $\pm$ 0.3 g cm$^{-3}$ for this comet.

\noindent
{\em Other works.} Rickman (1989) estimated M $\sim$ [1.3 - 3.1] $\times$ 10$^{14}$ kg. Later, Skorov and Rickman (1999) revised the Rickman's (1989) estimate and obtained $\rho$ = [0.5 - 1.2] g cm$^{-3}$, with
a preference for the lower value. Sagdeev {\em et al.} (1987) derived $M$ = 3 $\times$ 10$^{14}$ kg, and a bulk density of 0.6$^{+0.9}_{-0.6}$ g cm$^{-3}$, also based on a non-gravitational model. Peale (1989) estimated $\rho$ = [0.03 - 4.9] g cm$^{-3}$, with a preference for $\rho$ = 0.7 g cm$^{-3}$. These results are consistent with ours.

\subsubsection{2P/Encke}

{\em Mass.} We derive a mass of [9.2 $\pm$ 5.8] $\times$ 10$^{13}$ kg for this comet.

\noindent
{\em Size and density.} Encke has the shortest orbital period among the known comets. It is also a NEC (i.e. a {\em Near Earth Comet}). As a consequence of its short periodicity, proximity to Earth, and orbital stability, this comet has been extensively studied from ground-based observations, and hence has several size determinations, also with different methods. Luu and Jewitt (1990) estimate an effective nucleus radius $R_{Neff}$ of 3.28 $\pm$ 0.06 km from near-aphelion and time series CCD photometry. Lowry and Weissman (2007) determine $R_{Neff}$ = 3.95 $\pm$ 0.06 km by using the same technique. From radar and IR study Nolan and Harmon (2005) estimate $R_{Neff}$ = 2.42 $^{+0.86}_{-0.06}$ km. Tancredi {\em et al.} (2006) derive $R_{Neff}$ = 1.95 $^{+1.96}_{-0.67}$ km from the estimated absolute nuclear magnitude. Since there are for this comet several different estimates of the nucleus size, for which some of them yield unphysical values of $\rho$ when the corresponding errors are considered, we preferred to assume a sort of averaged radius, with a certain error estimate. After analyzing the different estimates, we adopted $R_{Neff}$ $\sim$ 3 $\pm$ 1 km. The corresponding bulk density would be then $\rho$ = 0.8 $\pm$ 0.8 g cm$^{-3}$.

\noindent
{\em Other works.} Rickman {\em et al.} (1987) estimated $M$ $\sim$ [2.4 - 3.2] $\times$ 10$^{13}$ kg, which is about half the value we obtained. We find that the cause of this discrepancy could be that Rickman {\em et al.} considered a  non-gravitational $\Delta$P that was a factor of 2 larger than ours.

\subsubsection{6P/d'Arrest}

{\em Mass.} We derive a mass of [2.8 $\pm$ 0.8] $\times$ 10$^{12}$ kg for this comet.

\noindent
{\em Size and density.} Tancredi {\em et al.} (2006) estimate $R_{Neff}$ = 1.7 $\pm$ 0.1 km. According to this value for the comet size we derive a bulk density of 0.15 $\pm$ 0.05 g cm$^{-3}$ for this comet.

\noindent
{\em Other works.} Szutowicz and Rickman (2006) estimate $M$ = 7.0 $\times$ 10$^{12}$ kg, from non-gravitational modeling. This value is larger than ours by a factor of 2.7. The lightcurve obtained and the $\Delta$P considered by these authors are similar to ours, so we looked for differences in the values assumed for the model parameters. Szutowicz and Rickman determine an empirical value for the lag angle $\eta$ by fitting an asymmetric model to astrometric observations performed between 1851 and 2001. They obtained $\eta$ $\sim$ 14$^{\circ}$, which agrees very well with our assumption. On the other hand, they assume a mean effective outflow velocity $<u>$ = 0.6 km s$^{-1}$, which is larger than the value we assume by a factor of 2.2. This would imply a very collimated flux with a very high momentum transfer efficiency.  Since the mass is directly proportional to this parameter, this explains the difference between both mass estimates. Szutowicz and Rickman also used different calibration coefficients: they assumed the value given by Jorda {\em et al.} (1992) for the slope coefficient ($a_1$ = -0.240), and determine $a_2$ = 30.5 by fitting the observations to measured gas production rates. We conclude that the different values assumed for the mean effective outflow velocity, as well as for the calibration coefficients, account for the discrepancy between both results. In principle, Szutowicz and Rickman's adopted value of $<u>$ seems to be extremely high.

\subsubsection{9P/Tempel 1}

{\em Mass.} We derive a mass of [2.3 $\pm$ 1.6] $\times$ 10$^{13}$ kg for this comet.

\noindent
{\em Size and density.} A'Hearn {\em et al.} (2005) estimate $R_{Neff}$ = 3.0 $\pm$ 0.1 km from close-up images taken by the {\em Deep Impact} fly-by spacecraft. According to this value for the comet size we derive a bulk density of 0.2 $\pm$ 0.1 g cm$^{-3}$ for this comet.

\noindent
{\em Other works.} Richardson and Melosh (2006) obtain $M$ = 5.0 $\times$ 10$^{13}$ kg by studying the expansion of the base of the conical ejecta plume  created by the impact of a 370 kg projectile delivered by the {\em Deep Impact} spacecraft on the comet nucleus.  This is the most direct measurement obtained to date (and {\em independent of the non-gravitational effect}, as we would like to emphasize). They obtain a bulk density $\rho$ = 0.4 $\pm$ 0.3 g cm$^{-3}$. Davidsson {\em et al.} (2007) obtain $M$ = (5.8 $\pm$ 1.6) $\times$ 10$^{13}$ kg, and $\rho$ = 0.45 $\pm$ 0.25 g cm$^{-3}$, by utilizing a sophisticated thermophysical model of the non-gravitational forces (which in addition relies on the measured gas production rates instead of the lightcurve). Our results are consistent with both independent works, which gives support to the usefulness of the modeling method used here.

\subsubsection{10P/Tempel 2}

{\em Mass.} We derive a mass of [3.5 $\pm$ 1.5] $\times$ 10$^{14}$ kg for this comet.

\noindent
{\em Size and density.} Jewitt and Luu (1989) obtain $R_{Neff}$ = 5.3 $^{+0.2}_{-0.7}$ km from time series CCD photometry. Tancredi {\em et al.} (2006) estimate $R_{Neff}$ = 4.02 $^{+1.49}_{-0.96}$ km from the estimated absolute nuclear magnitude. From the different estimates we adopted an averaged value $R_{Neff}$ $\sim$ 4.8 $\pm$ 0.7 km. According to this value for the comet size, we derive a bulk density of 0.7 $\pm$ 0.4 g cm$^{-3}$.

\noindent
{\em Other works.} Rickman {\em et al.} (1991) obtain $M$ = (1.6 $\pm$ 0.5)  $\times$ 10$^{14}$ kg, from non-gravitational force modeling. This is $\sim$ 50 \% of the value derived in this work. The main cause of this discrepancy could be that Rickman {\em et al.} computed the mass considering only the radial non-gravitational force component, and hence neglecting the contribution from the transverse component (according to our results, the transverse component contributes to $\sim$ 50 \% of the mass estimate).From the rotational lightcurve Jewitt and Luu (1989) derived parameters $a/b$ and $P_{rot}$ (the axial ratio and the rotational period, respectively), and obtained a minimum bulk density $\rho_{min}$ by assuming a negligible tensile strength of the nucleus material. They calculated $\rho_{min}$ = 0.3 g cm$^{-3}$, which is consistent with our result.

\subsubsection{19P/Borrelly}

{\em Mass.} We derive a mass of [2.7 $\pm$ 2.1] $\times$ 10$^{12}$ kg for this comet.

\noindent
{\em Size and density.} From HST observations Lamy {\em et al.} (1998) obtain semi-axes $a$ = 4.4 km and $b$ = $c$ = 1.8 km by approximating the nucleus to a triaxial ellipsoid, which corresponds to $R_{Neff}$ = 2.4 $\pm$ 0.2 km. This was later confirmed by direct imaging of the nucleus by {\em Deep Space 1} (Soderblom {\em et al.} 2002). According to this value for the comet size we derive a bulk density of 0.05 $\pm$ 0.04 g cm$^{-3}$ for this comet.

\noindent
{\em Other works.} Davidsson and Guti\ee rrez (2004) constrain the mass within the range $M$ = [8 - 24] $\times$ 10$^{12}$ kg, and the bulk density in the range $\rho$ = [0.10 - 0.30] g cm$^{-3}$, based on a thermophysical modeling of the non-gravitational force. Rickman {\em et al.} (1987) derive an upper limit $M$ $<$ 8.4 $\times$ 10$^{12}$ kg, which is consistent with our result. Davidsson (2001) obtains a minimum bulk density $\rho_{min}$ = 0.04 g cm$^{-3}$ by requiring that the nucleus withstands disruption due to centrifugal forces, assuming a strengthless nucleus material. By comparison with these results, we may conclude that we somewhat underestimated the mass and bulk density of Borrelly. Nevertheless, all the results suggest that we are dealing with an extremely low density comet. In disagreement with these results, Farnham and Cochran (2002) obtain a mass estimate that is substantially higher: they derive $M$ = 3.3 $\times$ 10$^{13}$ kg and $\rho$ = 0.49 g cm$^{-3}$ by following the method introduced by Rickman {\em et al.} (1987), but using a different vaporization model for the water production. They observed a strong primary (sunward) jet that was aligned with the nucleus' spin axis (according to their derived pole position). They also assumed an relatively small active area (4 \% of the nucleus' surface area), close enough to the pole that it would be illuminated almost continuously during the rotation of the nucleus. The combination of these two features (spin orientation and location and size of the primary source of activity) will produce a strong collimation of the jet,  making the non-gravitational force much more efficient (e.g. the non-gravitational force would produce the same non-gravitational effect $\Delta$P acting on a larger nucleus mass, than a less efficient non-gravitational force acting on a smaller nucleus mass). They assumed $<u>$ = 0.33 km s$^{-1}$.

\subsubsection{22P/Kopff}

{\em Mass.} We derive a mass of [5.3 $\pm$ 2.2] $\times$ 10$^{12}$ kg for this comet.

\noindent
{\em Size and density.} Tancredi {\em et al.} (2006) estimate $R_{Neff}$ = 1.8 $\pm$ 0.2 km from the estimated absolute nuclear magnitude.  According to this value for the comet size we derive a bulk density of 0.2 $\pm$ 0.1 g cm$^{-3}$ for this comet.

\noindent
{\em Other works.} Rickman {\em et al.} (1987) derive an upper limit $M$ $<$ 2.9 $\times$ 10$^{13}$ kg, which is consistent with our result.

\subsubsection{45P/Honda-Mkros-Pajdusakova}

{\em Mass.} We derive a mass of 1.9 $\times$ 10$^{11}$ kg for this comet, with an upper limit of 5.4 $\times$ 10$^{11}$ kg. The lower limit turns out to be negative which is unphysical. Therefore the computed mass of this comet is very uncertain and should be taken with extreme caution. The main reason for this uncertainty is that the radial and transverse terms $I_r$, $I_t$ that intervene in the computation of the mass (cf. eq. \ref{masa-aprox}) turn out to be of similar absolute values but different signs (see Table 5 below), so small errors in $I_r$ and $I_t$ can lead to either positive or negative values in ($I_r$ + $I_t$).

\noindent
{\em Size and density.} Tancredi {\em et al.} (2006) estimate $R_{Neff}$ = 0.33 $\pm$ 0.07 km from the estimated absolute nuclear magnitude. This would be the smallest comet of the sample, and one of the smallest known. According to this value for the comet size we derive a bulk density of 1.2 g cm$^{-3}$ for this comet. But we must emphasize that the size estimate is very uncertain, according to the referred authors. Because of the large uncertainty (not only in the nuclear size but also in the computed mass) we decided to exclude this comet from the general discussion between masses and densities presented in Section 7.

\noindent
{\em Other works.} Rickman {\em et al.} (1987) derive  $M$ = [1.5 - 5.4] $\times$ 10$^{11}$ kg, which is consistent with our result.

\subsubsection{46P/Wirtanen}

{\em Mass.} We derive a mass of [3.3 $\pm$ 2.3] $\times$ 10$^{11}$ kg for this comet.

\noindent
{\em Size and density.} Tancredi {\em et al.} (2006) estimate $R_{Neff}$ = 0.58 $\pm$ 0.03 km from the estimated absolute nuclear magnitude. According to this value for the comet size we derive a bulk density of 0.4 $\pm$ 0.3 g cm$^{-3}$ for this comet.

\noindent
{\em Other works.} We have not found another mass estimate for this comet.

\subsubsection{67P/Churyumov-Gerasimenko}

{\em Mass.} We derive a mass of [1.5 $\pm$ 0.6] $\times$ 10$^{13}$ kg for this comet.

\noindent
{\em Size and density.} From HST observations Lamy {\em et al.} (2006) obtain $R_{Neff}$ = 1.98 $\pm$ 0.02 km. According to this value for the comet size we derive a bulk density of 0.5 $\pm$ 0.2 g cm$^{-3}$ for this comet.

\noindent
{\em Other works.} Davidsson and Guti\ee rrez (2005) obtain $M$ = [0.35 - 2.1] $\times$ 10$^{13}$ kg, and the bulk density in the range $\rho$ = [0.10 - 0.60] g cm$^{-3}$, based on a sophisticated thermophysical modeling of the non-gravitational force. Our result is consistent with their work.

\subsubsection{81P/Wild 2}

{\em Mass.} We derive a mass of 8.1 $\times$ 10$^{12}$ kg for this comet, with an upper limit of 2.5 $\times$ 10$^{13}$ kg. The lower limit turns out to be negative which is unphysical. Therefore the computed mass of this comet is very uncertain and should be taken with extreme caution. The main reason for this uncertainty is the same as for comet 45P/H-M-P.

\noindent
{\em Size and density.} Brownlee {\em et al.} (2004) and Duxbury {\em et al.} (2004) found that the comet can be approximated by a triaxial ellipsoid with semi-axes \{$a,b,c$\} = \{2.75,2.00,1.65\} $\pm$ 0.05 km, from {\em Stardust} images. This corresponds to $R_{Neff}$ = 2.09 $^{+0.05}_{-0.06}$ km.
Howington-Kraus {\em et al.} (2005) performed a refined image analysis obtaining \{$a,b,c$\} = \{2.607,2.002,1.350\} $\pm$ 0.001 km, which corresponds to $R_{Neff}$ = 1.917 km. According to this value for the comet size we derive a bulk density of 0.3 g cm$^{-3}$ with an upper limit of 0.8 g cm$^{-3}$.

\noindent
{\em Other works.} Davidsson and Guti\ee rrez (2006) determined an upper limit $M$ $\le$ 2.3 $\times$ 10$^{13}$ kg, and $\rho$ $\le$ 0.60 g cm$^{-3}$, based on a sophisticated thermophysical modeling of the non-gravitational force. Our result is consistent with their work.\\

In Table 5 we summarize our results, namely the components $I_r$, $I_t$, the computed masses and bulk densities, the latter based on the radii found in the literature, also included in the table.

\begin{table*}
 \centering
 \begin{minipage}{140mm}
\caption{ Masses and densities.}
\begin{small}
\begin{tabular}{l l l l l l l} \hline
Comet & $I_r$  & $I_t $  & {\bf M}  & $R_N$ & {\small Ref. \footnote{(a) Keller {\em et al.} (1987), (b) See Section 6.2.2, (c) Tancredi {\em et al.} (2006), (d) A'Hearn {\em et al.} (2005),
(e) See Section 6.2.5, (f) Lamy {\em et al.} (1998), (g) Lamy {\em et al.} (2006), (h) Howington-Krauss {\em et al.} (2005).}} & {\bf $\rho$}  \\
 & {\footnotesize (molec. AU$^{-1}$)} & {\footnotesize (molec. AU$^{-1}$)} & {\footnotesize (kg)} & {\footnotesize (km)} & & {\footnotesize (g cm$^{-3}$)}\\ \hline
{1P/Halley}   & +1.66$\times$10$^{36}$ & +1.33$\times$10$^{36}$ & 3.2$\times$$10^{14}$ & 5.2 &(a) & 0.5 \\
{2P/Encke }    & -8.34$\times$10$^{34}$ & -6.96$\times$10$^{34}$ & 9.2$\times$$10^{13}$ & 3 &(b) &  0.8 \\
{6P/d'Arrest}  & +2.56$\times$10$^{34}$ & +1.09$\times$10$^{34}$& 2.8$\times$$10^{12}$ & 1.7 &(c) & 0.15 \\
{9P/Tempel 1 } & -1.60$\times$10$^{33}$& +5.42$\times$10$^{33}$& 2.3$\times$$10^{13}$ & 3.0 &(d)& 0.2 \\
{10P/Tempel 2} & +3.31$\times$10$^{34}$ & +3.16$\times$10$^{34}$& 3.5$\times$$10^{14}$ & 4.8 &(e) & 0.7 \\
{19P/Borrelly} & +1.39$\times$10$^{34}$& -2.98$\times$10$^{34}$ & 2.7$\times$$10^{12}$ & 2.4 &(f)& 0.05 \\
{22P/Kopff }   & -1.55$\times$10$^{34}$ & -3.24$\times$10$^{34}$& 5.3$\times$$10^{12}$ & 1.8 &(c) & 0.2 \\
{45P/H-M-P} & +2.45$\times$10$^{34}$& -3.26$\times$10$^{34}$ & (1.9 $\times$$10^{11}$) & 0.33 &(c)& (1.2) \\
{46P/Wirtanen} & +8.26$\times$10$^{32}$& -1.24$\times$10$^{34}$& 3.3$\times$$10^{11}$ & 0.58 &(c)& 0.4 \\
{67P/C-G }     & +1.46$\times$10$^{34}$ & +6.40$\times$10$^{33}$ &1.5$\times$$10^{13}$ & 1.98 &(g)& 0.5 \\
{81P/Wild 2}   & -1.67$\times$10$^{34}$ & +2.28$\times$10$^{34}$&(8.1$\times$$10^{12}$) & 1.917 &(h)& (0.3) \\
\hline
\end{tabular}
\end{small}
\end{minipage}
\end{table*}

According to the results of Table 5, the range of cometary masses of our sample cover several orders of magnitude, ranging from $\sim$ 3-4$\times$10$^{14}$ kg (Halley and Tempel 2) down to $\sim$ 2-3$\times$10$^{11}$ kg (H-M-P, Wirtanen), while the densities are almost all of the same order (tenths of g cm$^{-3}$), and below $\sim$ 0.8 g cm$^{-3}$ with a mean value $\approx$ 0.4 g cm$^{-3}$ (we excluded comet H-M-P due to its larger uncertainty in the size, as compared to the rest of the comets). This result is consistent with the cut-off value  $\sim$ 0.6 g cm$^{-3}$ for cometary densities suggested by Lowry and Weissman (2003), Snodgrass {\em et al.} (2006).
In the case of Borrelly we obtain en extremely low density of 0.05 g cm$^{-3}$ (very close to the lower limit imposed by the centrifugal forces, according to Davidsson 2001) though the large uncertainty of this estimate admits values more in line with the rest of the comets. Thermophysical modeling provides a higher value for Borrelly's density, though still pointing to a very low-density comet.

\subsection{The correlation between the non-gravitational change $\Delta$P and the perihelion asymmetries of the lightcurves}

We can quantitatively evaluate the perihelion asymmetry of the gas production curves by introducing the following definition

\begin{center}
\begin{equation}
A  \ = \ Q^{-1}_{m} \int_{0}^{P} Q(t) \sin[f(t)] dt
\label{asimetria}
\end{equation}
\end{center}

\noindent
where $t$, $f$ and $P$ are defined as before, and $Q_m$ is a normalization factor arbitrarily chosen to be the maximum water production rate of the comet. According to this definition, $A$ is measured in days.
In Fig. 15 we plot the perihelion asymmetry against the time of the maximum observed brightness as derived from the lightcurves (we remind the reader that the time is referred to the time of perihelion passage). As it should be expected, we can observe in general a linear relationship between both parameters, which also show very similar values.

We find that for those comets which show a strong positive perihelion asymmetry (i.e. a notorious post-perihelion excess of gaseous activity), the radial $I_r$ contribution dominates, with the exception of Borrelly. These are the cases of comets Halley, d'Arrest, Tempel 2, and C-G (actually, in the case of Tempel 2 the radial dominance is very weak). Wirtanen shows the most symmetrical lightcurve of the sample, with a weak positive perihelion asymmetry. In this case the transverse term $I_t$ turns out to be the dominant one, as expected.
Comets Kopff and Wild 2 show a {\em strong negative perihelion asymmetry} (i.e. a notorious pre-perihelion excess of activity), and a {\em dominant transverse contribution}. Even though the strong asymmetric lightcurves cause large radial components in the latter two comets, their transverse contributions are actually large too (a factor $\sim$ 1.4 - 2 greater than the radial ones). Tempel 1 also shows a negative asymmetry though a very weak one, so the transverse term dominates. Encke shows an intermediate negative perihelion asymmetry, but a slight dominant radial contribution instead.

Festou {\em et al.} (1990) and Rickman {\em et al.} (1991b) studied the correlation between the perihelion asymmetries of the gas production curves of periodic comets and the non-gravitational perturbations of their orbital periods. In this regard they defined the {\em standard non-gravitational change} $\Delta P'$ as the non-gravitational effect in the orbital period that the comet would experience if its semimajor axis was 3.5 AU, which is given by

\begin{equation}
 \Delta P' \ = \ \Delta P \Big( \frac{a}{3.5} \Big)^{-\frac{5}{2}},
 \label{DeltaPP}
 \end{equation}

\noindent
where $a$ is the semimajor axis. $\Delta P'$ then represents the non-gravitational effect $\Delta P$ corrected for different semi-major axes. Their result was that, in general, the perihelion asymmetries give the dominant contribution to the non-gravitational effect of the orbital period. Actually they found a linear correlation between the perihelion asymmetry and $\Delta P'$. This is expected if the $I_r$ contribution dominates, but not otherwise.

In Fig. 15 we plot the perihelion asymmetry (as defined in eq. (\ref{asimetria}), which is not exactly the same definition given by Festou {\em et al.} 1990, but it should be equivalent) as a function of $\Delta P'$ for the comets of the sample. According to Festou {\em et al.} (1990) and Rickman {\em et al.} (1991b), it should be expected an increase of the perihelion asymmetry with the increasing non-gravitational $\Delta P'$ (in absolute value). As it can be seen in the plot (bottom panel), we did not find such a linear correlation. As shown there, for some comets the transverse component $I_t$ becomes dominant, which weakens or completely blurs any possible correlation between $\Delta P'$ and the radial component $I_r$.

\begin{figure*}
\centering
\begin{minipage}{140mm}
\includegraphics[width=1.0\textwidth]{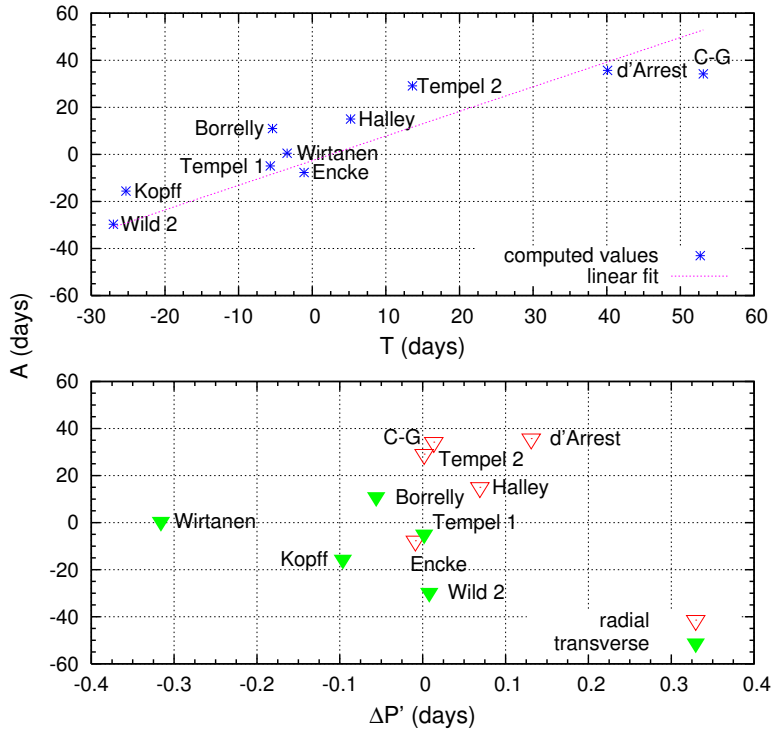}
\caption{{\bf Top}: Perihelion asymmetry $A$ as function of the time of the brightness maximum $T$ for the comets of the sample. A lineal fit to the computed values is also shown. {\bf Bottom}: Perihelion asymmetry $A$ as a function of the standard non-gravitational effect $\Delta P'$ for the comets of the sample. Those comets with a dominant contribution from the radial force component to the non-gravitational effect are marked with open triangles, and those comets with a dominant transverse force component contribution are marked with filled triangles.}
\end{minipage}
\end{figure*}

\section{Densities vs. masses in the Solar System}

We analyze the results obtained for our comet sample placing them into a broader context of different populations of bodies in the Solar System. In Fig. 16 we plot the bulk density as a function of the mass for the different populations of minor bodies in the Solar System as indicated in the figure caption. In the case of the comets we plot the computed masses and densities,  indicating the estimated errors for the computed masses and bulk densities as derived in Section 6.2. Fig. 16 shows that the computed comets form a more compact group than the other populations of minor bodies, and also present the lowest densities ($\rho$ $\lsim$ 0.8 g cm$^{-3}$). Our computed comets and the NEAs occupy the left of the diagram, corresponding to the smaller masses (except one NEA). The main distinction between both populations in this space of parameters is that the NEAs of the sample have larger bulk densities. Only two NEAs show $\rho$ $<$ 1 g cm$^{-3}$, but still above the upper limit of $\sim$ 0.8 g cm$^{-3}$ derived for the comets. Of course due to the smallness of the samples and the large uncertainties involved in the computation of masses and densities, both populations could actually somewhat overlap. This would be consistent with the idea that some NEAs could be deactivated JF comets (either extinct or dormant). Nevertheless, the clear separation in the density domain shown by comets and NEAs suggests that most NEAs are {\em bona fide} asteroids in agreement with Fern\'andez {\em et al.} (2002).

We find other objects not cataloged as comets but also with $\rho$ $<$ 1 g cm$^{-3}$; these are: Patroclus, Amalthea, 1999 TC$_{36}$, Antiope and Emma (the last two are MBAs). But in these cases the masses are at least about four orders of magnitude greater than the computed comet masses.

We note in Fig. 16 that no bodies with masses M $\gsim$ 10$^{20}$ kg have bulk densities $\lsim$ 1.5 g cm$^{-3}$. If we assume that comets and other ice-rich bodies of the outer solar system have similar compositions consisting of dust-ice mixtures, then the different densities would suggest that compaction of the material by self-gravity has taken place, so fluffy objects should not be expected for masses bigger than about 10$^{20}$ kg. We also have in the figure rocky bodies (NEAs and main-belt asteroids) whose differences in density may be due in this case to the different chemical composition, though different degrees of porosity might also be allowed. This is the case of most main-belt asteroids, TNOs, and, of course, the Moon and the largest Jovian satellites. Therefore, self-gravity would lead to noticeable variations in the primordial physical structure, ranging from loose aggregates to compacted and hardened material, and hence to a wide range of densities. In the case of the studied comets, due to the ranges of masses and sizes involved, we do no expect them to increase their densities by self gravity. Hence comets would be fluffy, low-density objects which have preserved their primordial fragile and porous structure.

\begin{figure*}
\centering
\begin{minipage}{140mm}
\includegraphics[width=1.0\textwidth]{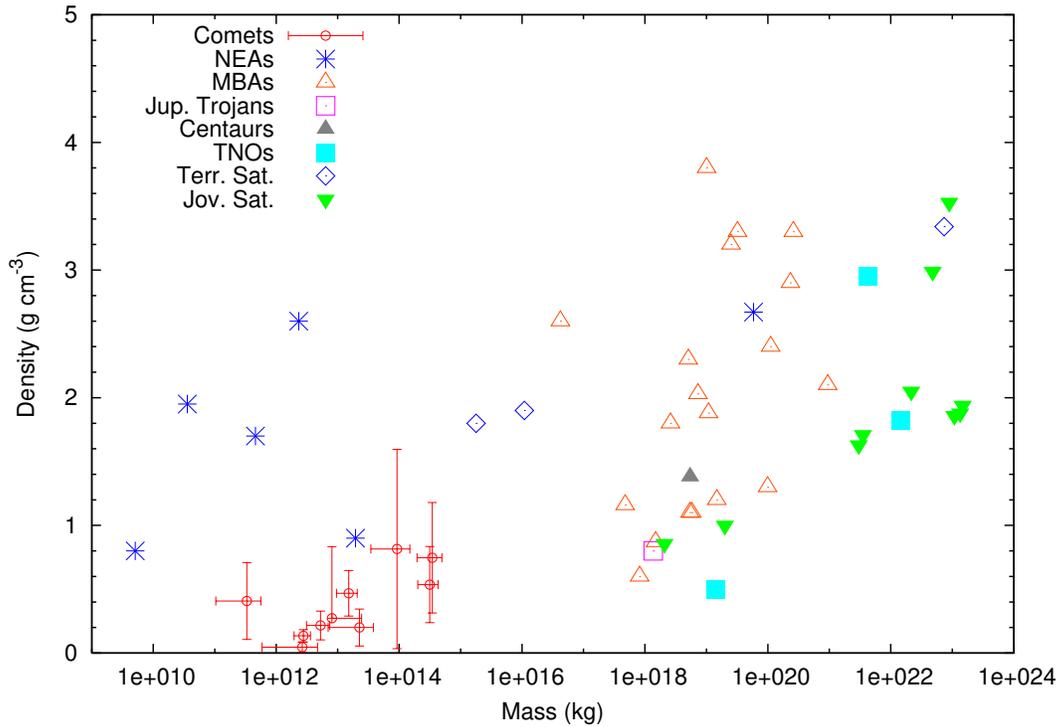}
\caption{Bulk density as a function of the mass for the different populations of minor bodies of the Solar System. For clarity of the figure, we only give error bars for our sample comets. The represented objects are: the ten comets of our sample (comet 45P/H-M-P is omitted due to its large uncertainty, and for comet 81P/Wild 2 only the nominal values and upper limits are given), nineteen objects from the Main Belt Asteroids (MBAs) population, six objects from the Near-Earth Asteroids (NEAs) population (Eros, Itokawa, 1999 KW$_4$, 2000 DP$_{107}$, 2000 UG$_{11}$, 2002 CE$_{26}$), one Trojan of Jupiter (Patroclus), one Centaur (Ceto-Phorcys), three trans-neptunian objects or TNOs (Pluto/Charon, 1999 TC$_{36}$, 2003 EL$_{61}$), three terrestrial satellites (Moon, Phobos and Deimos), and ten Jovian satellites. The mass and density (or size) data for these objects comes from: Yeomans {\em et al.} (2000) (433 Eros), Shinsuke {\em et al.} (2006) (Itokawa), Merline {\em et al.} (2002) (1999 KW4, 2000 DP$_{107}$, 2000 UG$_{11}$), Shepard {\em et al.} (2006) (2002 CE$_{26}$), Kova\u{c}evi\'{c} and Kuzmanoski (2007) (Ceres), Thomas {\em et al.} (2005) (size of Ceres and Pallas), Goffin (2001) (Pallas), Viateau and Rapaport (2001) (Vesta, Phartenope), Michalak (2001) (Hygiea, Eunomia), Lupishko (2006) (Psyche), Marchis {\em et al.} (2003) (Kalliope), Marchis {\em et al.} (2005b) (Eugenia, Camilla, Elektra, Emma, Huenna), Marchis {\em et al.} (2005a) (Sylvia), Marchis {\em et al.} (2005c) (Hermione), Descamps {\em et al.} (2005) (Antiope), Merline {\em et al.} (2002) (Ida, Pulcova), Marchis {\em et al.} (2006) (Patroclus), Grundy {\em et al} (2007) (Ceto-Phorcys), Tholen and Buie (1997) (Pluto/Charon), Stansberry {\em et al.} (2006) (1999 TC$^{36}$), Rabinowitz {\em et al.} (2006) (2003 EL$_{61}$), Anderson {\em et al.} (2005) (Amalthea), and references in De Pater and Lissauer (2001) (Moon, Phobos, Deimos, Amalthea, Io, Europe, Ganymedes, Calysto, Titan, Titania, Oberon, Triton, Nereida).
}
\end{minipage}
\end{figure*}

\section{Concluding remarks}

From non-gravitational force modeling we derive masses and densities for ten short-period comets of known sizes: 1P/Halley, 2P/Encke, 6P/d'Arrest, 9P/Tempel 1, 10P/Tempel 2, 19P/Borrelly, 22P/Kopff, 46P/Wirtanen, 67P/Churyumov-Gerasimenko and 81P/Wild 2. For another comet, 45P/H-M-P, it was not possible to compute a reliable mass and density. Our procedure follows the pioneer work of Rickman and colleagues (e.g. Rickman 1986, 1989; Rickman {\em et al.} 1987), and confirms the importance of the knowledge of non-gravitational effects on the comet's motion, as a fundamental tool to derive cometary masses. The present work is based on lightcurve data, the non-gravitational term $A_2$, and different assumptions for some physical parameters. Our main conclusions can be outlined as follows:

(i) The derived cometary masses cover more than three orders of magnitude, within the range $\sim$ [0.3 - 400] $\times$ 10$^{12}$ kg. Our best estimate for the bulk densities are below 0.8 g cm$^{-3}$, which is in agreement with models of the comet nucleus depicting it as a very low density object. The mean bulk density of our sample is 0.4 g cm$^{-3}$.

(ii) The results obtained are in general consistent with other works. Particularly, we remark the consistency with results from sophisticated thermophysical non-gravitational force modeling, and from the {\em Deep Impact} team using a different method for Tempel 1. In particular, this agreement gives us some reassurance of the usefulness of the non-gravitational method used here for deriving cometary masses, despite all the uncertainties inherent to it.

(iii) The method also relies on the empirical calibration between visual magnitudes and water production rates, which, in spite of its poor physical understanding, it has proved to be a very useful tool. Since there is an ample and comprehensive database of total magnitudes, the method could be extended to many more periodic comets, for which the non-gravitational change $\Delta P$ has been computed. If we also know comet sizes (whose knowledge is also rapidly improving) we can derive the comet's bulk density. Even though the computed bulk density for a given comet has a large uncertainty, by computing them for a large sample of comets we could narrow down the uncertainty of the whole sample to meaningful values.

(iv) In connection with the previous points, we want to stress that our results are meaningful from a statistical sense, by considering an \emph{average} value of a large sample of comets. Of course, this later statement is true if there are not systematic errors in the method by overestimating or underestimating any of the parameters used in the computation of the masses. Even tough this possibility cannot be ruled out, we have been very careful in choosing the values of the input parameters within physically realistic ranges. We should warn, however, that, given the large errors bars, we cannot tell at this point whether there is a real dispersion in the bulk densities of comets reflecting different degrees of compaction of the material and/or differences in the chemical composition.

(v) We conclude that a strong perihelion asymmetry of the lightcurve should not be necessarily related to a negligible transverse force component contribution to the non-gravitational change in the orbital period. Actually, in some cases (like comets Kopff and Wild 2), the transverse component is found to be more important than the radial contribution.

(vi) {\em Limitations of the method}. The mass solutions from eq. (\ref{masa-aprox}) are quite robust to small changes in the input parameters except when the terms $I_r$ and $I_t$ turn out to be of similar absolute value but different signs (as in the cases of 45P/H-M-P and 81P/Wild 2).

(vii) {\em Sources of uncertainty}. The lightcurve uncertainty and the values assumed for the model parameters, particularly for the effective outflow velocity $<u>$, would be in general the most important contribution to the uncertainty in the computed masses, although in some cases the non-gravitational change $\Delta P$ could become a major source of uncertainty too. An underestimate or overestimate of, e.g. $<u>$, would lead to a systematic underestimate or overestimate of the computed mass. A more detailed study of this parameter, like its dependance on the heliocentric distance, has to wait for more elaborated results from thermophysical models as well as more observations, particularly {\em in situ}.

\section*{Acknowledgments}

   We thank Pedro J. Guti\ee rrez for helpful discussions on this work and for providing data on gas production rates, and the referee for valuable comments and criticisms that helped to improve the presentation of the results. We also thank Daniel Green for providing data on electronic form from the ICQ. Andrea Sosa acknowledges financial support from the {\em Proyecto de Desarrollo de las Ciencias B\'asicas} (PEDECIBA) program to her MSc. thesis performed at {\em Universidad de la Rep\uu blica} of Uruguay. The present work has been performed as the main part of her thesis work.

%\section*{References}

\end{document}